\documentclass[superscriptaddress,amsmath,amssymb,prd,preprintnumbers,showpacs,twocolumn]{revtex4-2}

\usepackage{graphicx}
\usepackage{tabularx}
\usepackage[T1]{fontenc} 
\usepackage{amssymb,amsmath,bm,natbib}
\usepackage{color}
\usepackage{makecell}
\usepackage{slashed}
\usepackage{multirow}
\usepackage{graphics}
\usepackage[utf8]{inputenc}
\usepackage[caption=false]{subfig}
\usepackage{hyperref}
\usepackage{array}
\usepackage{url}
\usepackage{dsfont}
\usepackage{float}
\usepackage{cancel}
\usepackage{units}
\usepackage{blindtext}
\usepackage{upgreek}
\usepackage{booktabs}
\usepackage[dvipsnames,table,xcdraw]{xcolor}
\usepackage{enumerate}
\usepackage{mathtools}
\usepackage{soul,color}
\usepackage[normalem]{ulem}
\usepackage[vcentermath]{youngtab}

\newcommand{\orcid}[1]{\href{https://orcid.org/#1}{\includegraphics[width=10pt]{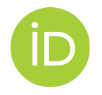}}}

\graphicspath{{./figures/pdf/}}

\begin{document}

\title{Crystallography, Lorentz violation, and the Standard-Model Extension}

\author{Marco Schreck\orcid{0000-0001-6585-4144}}
\affiliation{Programa de P\'{o}s-graduaç\~{a}o em F\'{i}sica, Universidade Federal do Maranh\~{a}o, Campus Universit\'{a}rio do Bacanga, S\~ao Lu\'is (MA), 65080-805, Brazil}
\affiliation{Coordena\c{c}\~{a}o do Curso de F\'{i}sica -- Bacharelado, Universidade Federal do Maranh\~ao, Campus Universit\'{a}rio do Bacanga, S\~ao Lu\'is (MA), 65080-805, Brazil}
\author{Rogeres A. da Silva Magalhães\orcid{0009-0003-0326-1943}}
\affiliation{Programa de P\'{o}s-graduaç\~{a}o em F\'{i}sica, Universidade Federal do Maranh\~{a}o, Campus Universit\'{a}rio do Bacanga, S\~ao Lu\'is (MA), 65080-805, Brazil}

\begin{abstract}

The motivation behind the present work is to adopt methodology from field theory and high-energy physics to crystallography. In particular, we establish a relationship between the electromagnetic sector of the Standard-Model Extension (SME) for Lorentz invariance violation and optical media. At an effective level, electromagnetic properties associated with different crystal structures are demonstrated to be parametrized in the SME. Crystallographic and magnetic point groups provide the mathematical tools to show this correspondence. Birefringent and magnetoelectric media merit a dedicated study. Intriguing effects, which have not been described systematically in the modern literature, are rediscovered for the latter and expressed in SME language. With the setting developed at our disposal, materials with specific symmetries such as birefringent or multiferroic crystals serve as condensed-matter analogs for SME effects. It enables us to propose materials with unusual optical properties, which have not been thoroughly looked at in recent times.

\end{abstract}


\maketitle

\section{Introduction}
\label{sec:introduction}

The main goal of this article is to demonstrate the usefulness of the Standard-Model Extension (SME) for condensed-matter and solid-state research on electromagnetic properties of materials. The SME is a generic relativistic field theory framework \cite{Colladay:1996iz,Colladay:1998fq} used especially in high-energy physics research for assessing experimental tests of spacetime symmetry violation. It extends the field theory of the Standard Model of elementary particles by terms that rest upon gauge and coordinate invariance but break Lorentz invariance, or CPT invariance, or both~\cite{Greenberg:2002uu}. Spacetime symmetry breaking is parametrized in terms of components of tensor-valued background fields, which are known as SME coefficients. The background fields are effective descriptions of inherent nontrivial vacuum properties caused by Planck-scale physics, such as strings \cite{Kostelecky:1988zi,Kostelecky:1991ak} or a small-scale spacetime structure, which loop quantum gravity~\cite{Gambini:1998it,Alfaro:2001rb,Bojowald:2004bb} and spacetime foam models~\cite{Bernadotte:2006ya,Hossenfelder:2014hha,Li:2023wlo} give rise to.


The SME has been under development since 1998. Kosteleck\'{y} and collaborators have extended the initial minimal framework on several occasions to include field operators with higher derivatives \cite{Kostelecky:2009zp,Kostelecky:2011gq,Kostelecky:2013rta,Ding:2016lwt,Kostelecky:2018yfa,Kostelecky:2020hbb}. Furthermore, the community has created and sharpened an ever increasing number of tools for solving certain problems. The latter comprise tree-level methods in quantum field theory~\cite{Kostelecky:2000mm,Adam:2001kx,Adam:2001ma,Adam:2002rg,Altschul:2004xp,Kaufhold:2005vj,Kaufhold:2007qd,Casana:2009xs,Casana:2010nd,Klinkhamer:2010zs,Klinkhamer:2011ez,Schreck:2011ai,Cambiaso:2012vb,Schreck:2013gma,Colladay:2014dua,Casana:2014cqa,Reis:2016hzu,Colladay:2016rsf,Ferreira:2020wde}, nonperturbative approaches~\cite{DelCima:2009ta,Potting:2011yj,Santos:2014lfa,SSantos:2015mzs}, tools of perturbation theory such as modified Feynman rules and renormalization~\cite{Kostelecky:2001jc,Colladay:2006rk,Colladay:2007aj,Colladay:2009rb,Ferrero:2011yu,Cambiaso:2014eba}, the computation of final-particle phase spaces for processes~\cite{Klinkhamer:2008ky,Hohensee:2008xz,Diaz:2013wia,Diaz:2015hxa,Colladay:2016rmy,Schreck:2017isa,Colladay:2017qfr,Amram:2023jlc,Petrov:2025wey}, and quantum corrections~\cite{Chung:1998jv,Chung:1999pt,Perez-Victoria:1999erb,Perez-Victoria:2001csb,Jackiw:1999yp,Altschul:2003ce,Altschul:2004gs,Brito:2007uc,Gomes:2009ch,BaetaScarpelli:2013rmt,Ferrari:2018tps,Altschul:2019eip,Altschul:2022isc,Altschul:2023vnf}.

Each term of the SME that goes beyond the Standard Model is supposed to describe tiny Lorentz-violating modifications of standard physics \textit{in vacuo}, which are expected to arise effectively as Planck-scale phenomena. These alterations should be measurable in experiments of high precision, high energy, or long duration. After all, each SME background field itself is beyond experimental control and remains unaffected by boosts and rotations of any experiment. Thus, the form of the fundamental laws of physics is expected to change with energy or direction. Since any such fundamental spacetime symmetry violations have evaded detection so far, many SME coefficients have been tightly constrained via laboratory and astrophysics experiments~\cite{Kostelecky:2008ts}.

The research performed here is dedicated to embedding electromagnetic material effects into an SME setting. Electromagnetism in material media is a relativistic \textit{U}(1) gauge theory. Unlike electromagnetism \textit{in vacuo}, it is an effective theory that loses its predictive power on typical interatomic or intermolecular length scales. Although any atomic lattice breaks fundamental Lorentz invariance, some material properties are relativistic at an effective level. In such a case, the invariant velocity of the Lorentz group is not the speed of light \textit{in vacuo}, but, e.g., the speed of light $c_m$ in an optical medium or the Fermi velocity of electrons in a material. This phenomenon has been coined emergent Lorentz invariance in the contemporary literature~\cite{Burkov:2017rgl,Kostelecky:2021bsb}.

Emergent Lorentz symmetry governs electromagnetic-wave propagation in material media. The latter differs from fundamental Lorentz symmetry since the medium speed of light $c_m$ is an effective quantity. In many materials, $c_m$ may depend on direction, frequency, or even polarization. Thereupon, emergent Lorentz invariance is broken, but the description retains its relativistic features. Nontrivial permittivity and permeability tensors, as well as magnetoelectric couplings, then parametrize all effects related to a broken emergent Lorentz invariance.

For example, if a material exhibits a permittivity that is described by a scalar, the latter is usually frequency-dependent, as is the form of the Maxwell equations. Alternatively, if a material has an optical axis, which is the case for certain birefringent crystals, the laws of physics depend on direction and polarization. Deformations of relativistic dispersion relations in momentum space can be modeled by background fields at the level of an effective field theory such as the SME. Apart from being a framework for experimental tests of fundamental spacetime symmetries, in the following, the SME will be reinterpreted as a stage for parametrizing violations of emergent Lorentz symmetry in the electromagnetism of materials.

The electromagnetic sector of the SME~\cite{Colladay:1998fq,Kostelecky:2002hh,Bailey:2004na} modifies Maxwell theory \textit{in vacuo} by a CPT-odd and a CPT-even term, respectively. The former is governed by a dimensionful vector-valued background field with controlling coefficients $(k_{AF})_{\kappa}$. Its timelike and spacelike components violate parity P and time reversal invariance T, respectively, whereas charge conjugation symmetry C is preserved. The latter contribution is parametrized by a dimensionless four-tensor of rank 4 with components $(k_F)_{\mu\nu\varrho\sigma}$, whose subsets either break or conserve P and T simultaneously.

The Lorentz tensor $k_F$ shares the symmetries of the Riemann curvature tensor. These symmetries reduce the initial 256 components of $k_F$ to a mere 20 independent ones. In any study of fundamental Lorentz invariance, an additional Lorentz-invariant component is removed via the requirement that $(k_F)^{\mu\nu}_{\phantom{\mu\nu}\mu\nu}=0$. However, as we shall see, this component is actually needed in the context of material media. Both $k_{AF}$ and $k_F$ with $|k_{AF}|\simeq \mathcal{O}(1)$ and $|k_F|\simeq \mathcal{O}(1)$ are expected to be capable of describing a wide range of electromagnetic properties of a material within a relativistic setting. To do so, we will set up modified constitutive relations for media with linear or approximately linear response.

The reduction of components of $k_F$ via symmetry properties of the latter tells us that further symmetries must be imposed on $k_F$ to render it compatible with special crystal structures. These additional symmetries are of internal nature and express how atoms or molecules are organized in a regular crystal lattice. In other words, $k_F$ must contain information on the specific geometry of the underlying Bravais lattice that exhibits discrete translation symmetry. Therefore, both $k_{AF}$ and $k_F$ are expected to capture features of the local atomic structure that are responsible for material properties on macroscopic scales.

At this juncture, we best resort to crystallographic groups, which provide an excellent mathematical tool to describe the internal symmetry properties of the atomic lattice making up the crystal at the microscopic level~\cite{Newnham:2005,Bilbao:2025,GraefMcHenry:2007}. We intend to establish a correspondence between crystallographic symmetries and configurations of the background field $k_F$. Our expectation is that each independent component is associated with a crystal lattice of specific symmetries. By doing so, we will systematically map different tensor components to electromagnetic material parameters. In particular, the known 32 crystallographic point groups classify component configurations of $k_F$ associated with distinctive material features.

A description of materials within a generic framework is expected to support the design of tailor-made materials according to the requirements of electronic devices. Compatibility between certain discrete symmetries and electromagnetism shapes electromagnetic properties, such as permittivity and permeability tensors or magnetoelectric responses. For example, if a design demands that a material have a suitable magnetoelectric coupling, possibly with anisotropies, the results provided here may serve as a guideline for identifying suitable compounds based on first principles. On the other hand, if a material has already been synthesized in the laboratory, the setup to be developed can help predict its electromagnetic characteristics to some extent. Last but not least, even the SME community might benefit from the findings to be made. After all, once an analogous condensed-matter system is available as an experimental sample, SME predictions can be tested in the laboratory.

The electrodynamics of exotic materials with nontrivial permittivity, permeability, and magnetoelectric tensors has been studied in a series of papers; see Refs.~\cite{Obukhov:2000nw,Obukhov:2004zz,Hehl:2007jy,Hehl:2007ut,Baekler:2014kha,Favaro:2014lja,Favaro:2015jxa,Hehl:2016wwp}. The authors of the latter articles do not employ the language of the SME. Our analysis has partial overlap with theirs, and we have been able to reproduce some of their results. Other than that, we will shed light on corners that the researchers have not explored in their manuscripts. Note that previous works of ours are dedicated to the electromagnetism of planar systems \cite{Ferreira:2019ygi,Lisboa-Santos:2023pwc}. In what follows, we will be working in a setting of three spatial dimensions.

Our manuscript is organized as follows. Section~\ref{sec:analogy} introduces the minimal electromagnetic SME and reviews its properties that are relevant to us. Section~\ref{eq:crystallographics-groups} is dedicated to crystallographic and magnetic point groups. Here, we will discuss which SME coefficients are compatible with the different crystal structures described by these groups. A particular interest of ours is to understand birefringence in the SME at all orders in the coefficients, which is explored in Sec.~\ref{eq:birefringence}. We conclude our findings in Sec.~\ref{eq:conclusions}. Appendix~\ref{app:formulas} is reserved for generic results that are important to state but are too long for the main body of the article. We use Heaviside-Lorentz units with $c=1$ unless otherwise stated. 

\section{Electrodynamics, Lorentz violation, and material media}
\label{sec:analogy}

The close of the second millennium witnessed a resurgence of research on Lorentz violation with the conception of the SME. The theory considered is a modification of Maxwell electrodynamics \textit{in vacuo} described by the following field-theoretic Lagrange density~\cite{Colladay:1998fq,Kostelecky:2002hh,Bailey:2004na}:
\begin{subequations}
\begin{align}
\label{eq:lagrangian}
\mathcal{L}&=\mathcal{L}_{\mathrm{Max}}+\mathcal{L}_{\mathrm{modMax}}+\mathcal{L}_{\mathrm{CFJ}}\,, \displaybreak[0]\\[1ex]
\mathcal{L}_{\mathrm{Max}}&=-\frac{1}{4}F_{\mu\nu}F^{\mu\nu}\,, \displaybreak[0]\\[1ex]
\label{eq:lagrangian-modMax}
\mathcal{L}_{\mathrm{modMax}}&=-\frac{1}{4}(k_F)^{\mu\nu\varrho\sigma}F_{\mu\nu}F_{\varrho\sigma}\,, \displaybreak[0]\\[1ex]
\label{eq:lagrangian-CFJ}
\mathcal{L}_{\mathrm{CFJ}}&=\frac{1}{2}(k_{AF})^{\kappa}\varepsilon_{\kappa\lambda\mu\nu}A^{\lambda}F^{\mu\nu}\,.
\end{align}
\end{subequations}
Here, $F_{\mu\nu}=\partial_{\mu}A_{\nu}-\partial_{\nu}A_{\mu}$ is the electromagnetic field strength tensor expressed in terms of the four-potential $A^{\mu}$, where $A^0=\phi$ is the electrostatic potential and $A^i$ are the components of the vector potential. All fields live in Minkowski spacetime with metric $\eta_{\mu\nu}$ of signature $(+,-,-,-)$. We also employ the four-dimensional totally antisymmetric Levi-Civita symbol $\varepsilon_{\kappa\lambda\mu\nu}$.

The first term in Eq.~\eqref{eq:lagrangian} is the Maxwell term \textit{in vacuo}. The second embodies a modification that is invariant under the combination CPT of charge conjugation, parity, and time reversal transformations.  The latter, which is sometimes called the modified Maxwell term, depends on a background field $k_F$, which transforms as a Lorentz tensor of rank 4 under rotations and boosts of the coordinate system. On the contrary, $k_F$ is fixed under physical rotations and boosts of an experiment. Therefore, in contrast to the dynamical electromagnetic fields $\mathbf{E}$ and $\mathbf{B}$, respectively, $k_F$ is interpreted as a nondynamical background field that is beyond the experimenter's control. It permeates the entire vacuum and is an effective property of the latter.

The background field $k_F$ inherits the symmetries of the bilinear combination $F_{\mu\nu}F_{\varrho\sigma}$ such that
\begin{subequations}
\begin{align}
(k_F)^{\mu\nu\varrho\sigma}&=-(k_F)^{\nu\mu\varrho\sigma}\,, \displaybreak[0]\\[1ex]
(k_F)^{\mu\nu\varrho\sigma}&=-(k_F)^{\mu\nu\sigma\varrho}\,, \displaybreak[0]\\[1ex]
(k_F)^{\mu\nu\varrho\sigma}&=(k_F)^{\varrho\sigma\mu\nu}\,.
\end{align}
\end{subequations}
When dealing with fundamental Lorentz violation, the completely antisymmetric piece of $k_F$ is eliminated, as it leads to a surface term. To do so, the background is required to satisfy the Bianchi-type identity
\begin{equation}
\label{eq:bianchi-type-identity}
(k_F)^{\mu\nu\varrho\sigma}+(k_F)^{\mu\varrho\sigma\nu}+(k_F)^{\mu\sigma\nu\varrho}=0\,.
\end{equation}
All in all, $k_F$ shares its symmetries with those of the Riemann tensor. Moreover, the double trace of $k_F$ can be absorbed into the Maxwell term by redefining the physical fields. Hence, the condition $(k_F)^{\mu\nu}_{\phantom{\mu\nu}\mu\nu}=0$ is usually imposed. The modified Maxwell term has been investigated to a certain extent, ranging from phenomenology in astroparticles \cite{Klinkhamer:2007ak,Klinkhamer:2008ky,Klinkhamer:2008ss,Schreck:2013paa,Diaz:2016dpk,Klinkhamer:2017puj,Duenkel:2021gkq,Duenkel:2021szq,Duenkel:2023nlk} to its quantum field theoretic properties \cite{Klinkhamer:2010zs,Klinkhamer:2011ez,Schreck:2011ai,Cambiaso:2012vb,Schreck:2013gma}.

Last but not least, the third term of Eq.~\eqref{eq:lagrangian} is noninvariant under CPT. It is known as the Carroll-Field-Jackiw (CFJ) term, named after the authors of the first paper~\cite{Carroll:1989vb} on the phenomenological consequences of this term in astrophysics. In principle, this contribution is of Chern-Simons type, but it is not a genuine, i.e., topological Chern-Simons term \cite{Chern:1974}; see also Ref.~\cite{Lisboa-Santos:2023pwc}. In fact, the latter involves a vector-valued background field denoted as $k_{AF}$. The CFJ term has been studied extensively in the contemporary literature; see, e.g., Refs.~\cite{Chung:1998jv,Chung:1999pt,Perez-Victoria:1999erb,Perez-Victoria:2001csb,Altschul:2003ce,Altschul:2004gs,Adam:2001kx,Adam:2001ma,Adam:2002rg,Kaufhold:2005vj,Brito:2007uc,Kaufhold:2007qd,Colladay:2016rmy,Altschul:2019eip,Ferreira:2020wde}.

The SME allows for describing certain effective properties of material media in the relativistic setting of classical electromagnetism. Some papers already highlight the applicability of terms contained in the SME to condensed-matter systems, such as topological semimetals; see Refs.~\cite{Grushin:2012mt,Zyuzin:2012tv,Landsteiner:2013sja,Miransky:2015ava,Armitage:2017cjs,Behrends:2018qkj,Kostelecky:2021bsb,Kostelecky:2025zsy}. When it comes to the electromagnetic sector of the SME, the symmetry-violating terms parametrized by $k_F$ and $k_{AF}$ can be mapped onto constitutive relations of anisotropic or magnetoelectric crystals. In this context, the components of $k_F$ and $k_{AF}$ contribute to the electric permittivity $\epsilon_{ij}$, magnetic permeability $\mu_{ij}$, and magnetoelectric couplings $\alpha_{ij}$, which are first and foremost taken as properties of a Lorentz-violating vacuum. The viewpoint taken in this paper is to replace the vacuum by a material, which turns $k_{AF}$ and $k_F$ into means to parametrize the electromagnetic features of this very material. In other words, Maxwell electrodynamics in material media shall be parametrized in terms of SME coefficients, i.e., it will be treated as a modification of Maxwell theory \textit{in vacuo}.

The 32 crystallographic and 122 magnetic point groups provide the mathematical tools to classify effective electromagnetic characteristics of a crystal at macroscopic scales according to its lattice at microscopic scales. For instance, materials with specific symmetries, e.g., birefringent or multiferroic crystals, serve as condensed-matter analogs for SME background field configurations. Choices of $k_F$ are able to replicate electromagnetic signatures due to anisotropies in real crystals. In principle, SME predictions can be tested in such materials.

For an introduction to crystallographic groups, the reader may consult Refs.~\cite{Newnham:2005,GraefMcHenry:2007}, if they so wish. We will be using the Hermann-Mauguin notation, where each group is named by a sequence of boldface numbers $\mathbf{1}$, $\mathbf{2}$, $\mathbf{3}$, $\mathbf{4}$, and $\mathbf{6}$. The latter indicate the presence of $n$-fold rotations, where $n\in\{1,2,3,4,6\}$ are the only possibilities according to a theorem. Note that a 1-fold rotation corresponds to the identity. The symbol $\mathbf{m}$ (``mirror'') is employed whenever a two-dimensional parity transformation (reflection at a plane) occurs. A bar placed on top of a boldface number means that a three-dimensional parity transformation (inversion at the coordinate origin) P is involved. The 7 Curie groups are the only ones that incorporate continuous ($\infty$-fold) rotations around certain axes. The symbol $\boldsymbol{\infty}$ indicates that such transformations play a role. Finally, a prime (') only accompanies magnetic point groups and stands for an additional time reversal transformation T.

\subsection{Modified Maxwell theory}

First, we shall delve into an extended Maxwell electrodynamics that the modified Maxwell term of Eq.~\eqref{eq:lagrangian-modMax} gives rise to. For now, we discard the CFJ term of Eq.~\eqref{eq:lagrangian-CFJ}, which we will dedicate ourselves to later. The constitutive relations for homogeneous anisotropic media are expressed in SME notation through the following matrix equation~\cite{Kostelecky:2002hh}:
\begin{subequations}
\label{eq:constitutive-relations-SME}
\begin{equation}
\label{eq:matrix-equation}
\left(
\begin{array}
[c]{c}%
\mathbf{D}\\
\mathbf{H}%
\end{array}
\right)  =\left(
\begin{array}
[c]{cc}%
\epsilon & \alpha \\
-\alpha^{T} & \mu^{-1} \\
\end{array}
\right)  \left(
\begin{array}
[c]{c}%
\mathbf{E}\\
\mathbf{B}%
\end{array}
\right)\,, \\[1ex]
\end{equation}
with
\begin{align}
\label{eq:permittivity}
\epsilon&=\mathds{1}_3+\kappa_{DE}\,,\quad \mu^{-1}=\mathds{1}_3+\kappa_{HB}\,, \\[1ex]
\label{eq:definition-beta}
\alpha&=\kappa_{DB}\,,\quad \alpha^T=-\kappa_{HE}\,,
\end{align}
\label{eq:constitutive-relations-fields}%
\end{subequations}
where $\mathbf{D}$ is the electric displacement field, $\mathbf{H}$ the magnetic field, $\mathbf{E}$ the electric field, and $\mathbf{B}$ the magnetic flux density. Moreover, $\epsilon$ is the permittivity tensor, $\mu$ the permeability tensor, and $\alpha$ is known as the magnetoelectric-coupling tensor. The SME variables $\kappa_{DE}$, $\kappa_{HB}$, $\kappa_{DB}$, and $\kappa_{HE}$ are $(3\times3)$ matrices, which are defined in terms of component coefficients of $k_F$:
\begin{subequations}
\label{eq:definitions-kappa-matrices}
\begin{align}
\label{eq:definition-kappaDE}
(\kappa_{DE})^{jk}&:=-2(k_F)^{0j0k}\,, \\[1ex]
\label{eq:definition-kappaHB}
(\kappa_{HB})^{jk}&:=\frac{1}{2}\varepsilon^{jpq}\varepsilon^{krs}(k_F)^{pqrs}\,, \\[1ex]
\label{eq:definition-kappaDB}
(\kappa_{DB})^{jk}&=-(\kappa_{HE})^{kj}:=(k_F)^{0jpq}\varepsilon^{kpq}\,.
\end{align}
\end{subequations}
Nonzero coefficients of these matrices parametrize modifications of Maxwell electrodynamics \textit{in vacuo} that govern electromagnetism in material media. The $(3\times 3)$ blocks on the diagonal of Eq.~\eqref{eq:matrix-equation} represent effective permittivity and permeability properties, while the off-diagonal $(3\times 3)$ blocks describe magnetoelectric couplings if they are present.

The compact matrix equation~\eqref{eq:constitutive-relations-SME} completely encodes all anisotropic features that respect relativistic coordinate invariance and the \textit{U}(1) gauge symmetry of electromagnetism. The modifications are interpreted as effective material properties, where anisotropic dielectric and magnetic responses occur alongside unconventional magnetoelectric couplings. By using the quantities of Eq.~\eqref{eq:definitions-kappa-matrices}, the Maxwell Lagrange density \textit{in vacuo} extended by Eq.~\eqref{eq:lagrangian-modMax} can be expressed in terms of the physical fields~\cite{Kostelecky:2002hh}:
\begin{align}
\label{eq:lagrange-density-physical-fields}
\mathcal{L}_{\mathrm{Max}}+\mathcal{L}_{\mathrm{modMax}}&=\frac{1}{2}\mathbf{E}\cdot(\mathds{1}_3+\kappa_{DE})\cdot\mathbf{E} \notag \\
&\phantom{{}={}}-\frac{1}{2}\mathbf{B}\cdot(\mathds{1}_3+\kappa_{HB})\cdot\mathbf{B} \notag \\
&\phantom{{}={}}+\mathbf{E}\cdot\kappa_{DB}\cdot\mathbf{B}\,.
\end{align}
The following constitutive relations associate applied electromagnetic fields to material responses for a linear magnetoelectric medium:
\begin{equation}
\label{eq:constitutive-relations-response}
\left(
\begin{array}
[c]{c}%
\mathbf{P}\\
\mathbf{M}%
\end{array}
\right)  =\left(
\begin{array}[c]{cc}%
\chi^{P} & \beta \\
\beta^{T} & \chi^{M} \\%
\end{array}
\right)  \left(
\begin{array}
[c]{c}%
\mathbf{E}\\
\mathbf{H}%
\end{array}
\right)\,,
\end{equation}
with polarization $\mathbf{P}$ and magnetization $\mathbf{M}$. The electric and magnetic susceptibility tensors $\chi^{P}$ and $\chi^{M}$, respectively, characterize the response of the medium to electric and magnetic fields. The (alternative) magnetoelectric-coupling tensor $\beta$ describes any nontrivial coupling between electric and magnetic degrees of freedom, which may be present.

We would like to formulate Eq.~\eqref{eq:constitutive-relations-response} in SME notation. To do so, it is beneficial to recast Eq.~\eqref{eq:constitutive-relations-fields} into the following form:
\begin{equation}
\label{eq:constitutive-relations-reformulated}
\left(
\begin{array}
[c]{c}%
\mathbf{D}\\
\mathbf{B}%
\end{array}
\right)  =\left(
\begin{array}
[c]{cc}%
\epsilon+\beta\mu\alpha^{T} & \alpha\mu \\
\mu\alpha^{T} & \mu%
\end{array}
\right)  \left(
\begin{array}
[c]{c}%
\mathbf{E}\\
\mathbf{H}%
\end{array}
\right)\,.
\end{equation}
Now, from the basic definitions
\begin{equation}
\mathbf{D}=\mathbf{E}+\mathbf{P}\,,\quad \mathbf{B}=\mathbf{H}+\mathbf{M}\,,
\end{equation}
and Eqs.~\eqref{eq:constitutive-relations-fields}, \eqref{eq:constitutive-relations-reformulated}, we arrive at
\begin{subequations}
\label{eq:susceptibilities}
\begin{equation}
\begin{pmatrix}
\mathbf{P} \\
\mathbf{M} \\
\end{pmatrix}=\begin{pmatrix}
\mathbf{D} \\
\mathbf{B} \\
\end{pmatrix}-\begin{pmatrix}
\mathbf{E} \\
\mathbf{H} \\
\end{pmatrix}=\begin{pmatrix}
\chi^{P} & \beta \\
\beta^{T} & \chi^{M} \\
\end{pmatrix}\begin{pmatrix}
\mathbf{E} \\
\mathbf{H} \\
\end{pmatrix}\,,
\end{equation}
with
\begin{align}
\chi^P&=\mathbb{\kappa}_{DE}+\kappa_{DB}(\mathds{1}_3+\kappa_{HB})^{-1}\kappa_{DB}^{T}\,, \\[1ex]
\chi^M&=(\mathds{1}_3+\kappa_{HB})^{-1}-\mathds{1}_3\,, \\[1ex]
\label{eq:magnetoelectric-coupling}
\beta&=\kappa_{DB}(\mathds{1}_3+\kappa_{HB})^{-1}\,.
\end{align}
\end{subequations}
The latter relationships completely characterize the electromagnetic properties of the material medium and relate the SME formalism to electromagnetic effects in materials. As long as $\kappa_{DB}=0$, the quantities $\kappa_{DE}$ and $\kappa_{HB}$ are directly linked to the electric and magnetic susceptibilities.

\subsubsection{Two principal sectors}
\label{sec:principal-sectors}

The symmetries of $k_F$, as well as the tracelessness condition $(k_F)^{\mu\nu}_{\phantom{\mu\nu}\mu\nu}=0$, amount to 19 independent component coefficients. The latter set decomposes into two sectors with different properties. The first principal sector involves 9 coefficients and is covered by the following parametrization~\cite{Bailey:2004na,Altschul:2006zz}:
\begin{equation}
\label{eq:nonbirefringent-ansatz}
(k_F)^{\mu\nu\varrho\sigma}=\frac{1}{2}(\eta^{\mu\varrho}\tilde{\kappa}^{\nu\sigma}-\eta^{\mu\sigma}\tilde{\kappa}^{\nu\varrho}-\eta^{\nu\varrho}\tilde{\kappa}^{\mu\sigma}+\eta^{\nu\sigma}\tilde{\kappa}^{\mu\varrho})\,,
\end{equation}
where $\tilde{\kappa}^{\mu\nu}=:(k_F)_{\alpha}^{\phantom{\alpha}\mu\alpha\nu}$ is a symmetric and traceless $(4\times 4)$ matrix. The defining feature of this sector is that the associated coefficients do not imply birefringence at first order in the coefficients. Now, some coefficient choices do not exhibit birefringence at all. However, in general, birefringence occurs at second order in $k_F$. When dealing with fundamental Lorentz violation such that $|k_F|\ll 1$, birefringence is completely suppressed in any case, which is why the SME community refers to this set as the nonbirefringent sector.

However, when applying Eq.~\eqref{eq:lagrangian} to the condensed-matter context, the previous name is misleading since $|k_F|\simeq \mathcal{O}(1)$. Thus, even when birefringence occurs at second order in $k_F$, these effects are far from being suppressed and can even turn out to be large. Therefore, in condensed-matter physics, it is reasonable to refrain from using this terminology. In contrast, we will be referring to the coefficients of the first principal sector or, alternatively, to the Ricci-like coefficients, in light of the definition of $\tilde{\kappa}^{\mu\nu}$ under Eq.~\eqref{eq:nonbirefringent-ansatz}.

The component coefficients of $k_F$ in terms of $\tilde{\kappa}^{\mu\nu}$ are
\begin{equation}
\label{eq:nonbirefringent-coefficients}
\begin{pmatrix}
(k_F)^{2323} \\
(k_F)^{1313} \\
(k_F)^{1212} \\
(k_F)^{1323} \\
(k_F)^{1223} \\
(k_F)^{1213} \\
(k_F)^{0313} \\
(k_F)^{0323}\\
(k_F)^{0223} \\
\end{pmatrix}=\begin{pmatrix}
(k_F)^{0101} \\
(k_F)^{0202} \\
(k_F)^{0303} \\
-(k_F)^{0102} \\
(k_F)^{0103} \\
-(k_F)^{0203} \\
(k_F)^{0212} \\
-(k_F)^{0112} \\
(k_F)^{0113} \\
\end{pmatrix}=\frac{1}{2}\begin{pmatrix}
-\tilde{\kappa}^{00}+\tilde{\kappa}^{11} \\
-\tilde{\kappa}^{00}+\tilde{\kappa}^{22} \\
-(\tilde{\kappa}^{11}+\tilde{\kappa}^{22}) \\
-\tilde{\kappa}^{12} \\
\tilde{\kappa}^{13} \\
-\tilde{\kappa}^{23} \\
-\tilde{\kappa}^{01} \\
-\tilde{\kappa}^{02} \\
\tilde{\kappa}^{03} \\
\end{pmatrix}\,.
\end{equation}
For these 9 coefficients, the matrices of Eq.~\eqref{eq:definitions-kappa-matrices} have the form
\begin{subequations}
\label{eq:matrices-kappa-sector1}
\begin{align}
\label{eq:kappa-DE-sector1}
\kappa_{DE}  &  =\left(
\begin{array}
[c]{ccc}%
\tilde{\kappa}^{00}-\tilde{\kappa}^{11} & -\tilde{\kappa}^{12} &
-\tilde{\kappa}^{13}\\
-\tilde{\kappa}^{12} & \tilde{\kappa}^{00}-\tilde{\kappa}^{22} &
-\tilde{\kappa}^{23}\\
-\tilde{\kappa}^{13} & -\kappa^{23} & \tilde{\kappa}^{11}+\tilde{\kappa}^{22}%
\end{array}
\right)\,, \displaybreak[0]\\[1ex]
\label{eq:kappa-HB-sector1}
\kappa_{HB}&=-\kappa_{DE}\,, \displaybreak[0]\\[1ex]
\label{eq:kappa-DB-sector-1}
\kappa_{DB}  &  =\left(
\begin{array}
[c]{ccc}%
0 & -\tilde{\kappa}^{03} & \tilde{\kappa}^{02}\\
\tilde{\kappa}^{03} & 0 & -\tilde{\kappa}^{01}\\
-\tilde{\kappa}^{02} & \tilde{\kappa}^{01} & 0
\end{array}
\right)  =\kappa_{HE}\,.
\end{align}
\end{subequations}
In the present sector, according to Eq.~\eqref{eq:constitutive-relations-fields}, $\kappa_{DE}$ completely determines
$\epsilon$ and $\mu$. The matrices $\kappa_{DE}$ and $\kappa_{HB}$ are symmetric by construction, and each involves the same 6 independent coefficients. Furthermore, the matrices $\kappa_{DB}$ and $\kappa_{HE}$ are antisymmetric and traceless, which makes these objects $\mathfrak{so}(3)$-valued. They also correspond to each other, which is not necessarily the case for arbitrary configurations. Thus, each contains a further 3 independent coefficients of the first principal sector, which sums up to a total of 9, as expected.

The remaining 10 component coefficients of $k_F$ are part of the second principal sector. They imply electromagnetic birefringence at first order in $k_F$. These 10 coefficients will be denoted as $k^a$ for $a\in\{1\dots 10\}$, and we can choose them from the component coefficients of $k_F$ as follows:
\begin{equation}
\label{eq:birefringent-coefficients}
\begin{pmatrix}
(k_F)^{0213} \\
(k_F)^{0123} \\
(k_F)^{1313} \\
(k_F)^{1212} \\
(k_F)^{1323} \\
(k_F)^{1223} \\
(k_F)^{1213} \\
(k_F)^{0323} \\
(k_F)^{0223} \\
(k_F)^{0313} \\
\end{pmatrix}=\begin{pmatrix}
k^1 \\
k^2 \\
-k^3 \\
-k^4 \\
k^5 \\
-k^6 \\
k^7 \\
k^8 \\
-k^9 \\
-k^{10} \\
\end{pmatrix}\,,
\end{equation}
with the additional choices $(k_F)^{2323}=k^3+k^4$ to impose $(k_F)^{\mu\nu}_{\phantom{\mu\nu}\mu\nu}=0$ and $(k_F)^{0312}=k^1-k^2$ to satisfy Eq.~\eqref{eq:bianchi-type-identity}.
Note that there is a certain freedom in choosing these coefficients and alternative parametrizations can be conceived; see Ref.~\cite{Kostelecky:2002hh}. Our motivation for Eq.~\eqref{eq:birefringent-coefficients} is its simplicity.

For $\tilde{\kappa}^{\mu\nu}=0$, the matrices of Eq.~\eqref{eq:definitions-kappa-matrices} are expressed as
\begin{subequations}
\label{eq:matrices-kappa-sector2}
\begin{align}
\label{eq:kappa-DE-sector2}
\kappa_{DE}  &  =\left(
\begin{array}
[c]{ccc}%
0 & 0 & 0\\
0 & 0 & 0\\
0 & 0 & 0
\end{array}
\right)\,, \displaybreak[0]\\[1ex]
\label{eq:kappa-HB-sector2}
\kappa_{HB}&  =-2\left(
\begin{array}
[c]{ccc}%
-(k^3+k^4) & k^{5} & k^{6}\\
k^{5} & k^{3} & k^{7} \\
k^{6} & k^{7} & k^{4} \\
\end{array}
\right)\,, \displaybreak[0]\\[1ex]
\label{eq:kappa-DB-sector-2}
\kappa_{DB}  &  =2\left(
\begin{array}
[c]{ccc}%
k^{2} & 0 & 0\\
-k^{9} & -k^{1} & 0\\
k^{8} & k^{10} & k^{1}-k^{2}%
\end{array}
\right)=-(\kappa_{HE})^T\,.
\end{align}
\end{subequations}
Interestingly, none of the coefficients contained in the second principal sector contributes to $\kappa_{DE}$. Moreover, neither $\kappa_{DB}$ nor $\kappa_{HE}$ are now antisymmetric matrices. Considering all 19 coefficients of both sectors simultaneously simply means that the matrices of the first sector, Eq.~\eqref{eq:matrices-kappa-sector1}, and the corresponding ones of the second sector, Eq.~\eqref{eq:matrices-kappa-sector2}, have to be added.
\begin{table}
\centering
\begin{tabular}{cccc}
\toprule
Coefficient & C & P & T \\
\midrule
$\kappa^{00},\kappa^{ij},k^3,k^4,k^5,k^6,k^7$ & $+$ & $+$ & $+$ \\
$\kappa^{0i},k^1,k^2,k^8,k^9,k^{10}$ & $+$ & $-$ & $-$ \\
\bottomrule
\end{tabular}
\caption{Transformation properties of coefficients for the two principal sectors of $k_F$ according to Eqs.~\eqref{eq:nonbirefringent-coefficients} and \eqref{eq:birefringent-coefficients} as well as Ref.~\cite{Kostelecky:2008ts}.}
\label{eq:CPT-behavior-coefficients-modMax}
\end{table}

According to Tab.~\ref{eq:CPT-behavior-coefficients-modMax}, the matrices $\kappa_{DE}$ and $\kappa_{HB}$ only contain T-even SME coefficients, i.e., these matrices are invariant under T. On the contrary, $\kappa_{DB}$ and $\kappa_{HE}$ involve the remaining T-odd coefficients. This makes sense, since the latter two matrices parametrize magnetoelectric couplings.

Note that 6 out of the 8 matrices of Eqs.~\eqref{eq:matrices-kappa-sector1} and~\eqref{eq:matrices-kappa-sector2} are traceless except $\kappa_{DE}$ and $\kappa_{HB}$ of the first principal sector, stated in Eqs.~\eqref{eq:kappa-DE-sector1} and \eqref{eq:kappa-HB-sector1}, respectively. Hence, potential conflicts emerge for materials that are characterized by either a trivial permittivity or permeability. To avoid such hurdles, the double tracelessness condition on $k_F$ should be dropped. Let us define a nonzero double trace of $k_F$ by $(k_F)^{\mu\nu}_{\phantom{\mu\nu}\mu\nu}:=k_{\mathrm{tr}}$, which complements the previous 19 independent coefficients of $k_F$ to 20. This additional coefficient is introduced by hand via the replacement rules $\kappa_{DE}\rightarrow \kappa_{DE}+(k_{\mathrm{tr}}/6)\mathds{1}_3$ and $\kappa_{HB}\rightarrow \kappa_{HB}+(k_{\mathrm{tr}}/6)\mathds{1}_3$. For the individual coefficients of the $k_F$ tensor, this means that $(k_F)^{0i0i}\rightarrow (k_F)^{0i0i}-k_{\mathrm{tr}}/12$ and $(k_F)^{ijij}\rightarrow (k_F)^{ijij}+k_{\mathrm{tr}}/12$ for $i\neq j$.

In studies of fundamental Lorentz violation \textit{in vacuo}, the latter double trace is discarded since it does not contribute to Lorentz-violating physics. As already stated and will become clear below, a nonzero $k_{\mathrm{tr}}$ may be indispensable when describing electromagnetism in material media by the SME.

\subsection{Carroll-Field-Jackiw term}
\label{eq:CFJ-theory}

In the following, the CFJ term of Eq.~\eqref{eq:lagrangian-CFJ} is switched on, whereas the modified Maxwell term of Eq.~\eqref{eq:lagrangian-modMax} shall be put to rest for now. In fact, it is possible to recast the CFJ contribution in terms of the dual electromagnetic tensor $\tilde{F}_{\mu\nu}$:
\begin{subequations}
\begin{align}
\mathcal{L}_{\mathrm{CFJ}}&=(k_{AF})^{\kappa}A^{\lambda}\tilde{F}_{\kappa\lambda}\,, \\[2ex]
\tilde{F}_{\mu\nu}&:=\frac{1}{2}\varepsilon_{\mu\nu\varrho\sigma}F^{\varrho\sigma}\,.
\end{align}
\end{subequations}
We note in passing that $(k_{AF})^0$ preserves C and T but violates P. Contrarily, $(k_{AF})^i$ preserves C and P but violates T. Furthermore, the CFJ term, as defined in Ref.~\cite{Carroll:1989vb}, differs from ours by a factor of $-2$. Interestingly, CFJ theory is related to deep physics developed in the 70s and 80s.

Chiral symmetry breaking at the quantum level in an SU($N$) gauge theory was initially observed by Adler, Bell, and Jackiw \cite{Adler:1969gk,Bell:1969ts}. The anomaly depends on $G^a_{\mu\nu}\tilde{G}^{a,\mu\nu}$, where $G_{\mu\nu}^a$ and $\tilde{G}^{a,\mu\nu}$ are the components of the $\mathfrak{su}(N)$-valued field strength tensor and of its dual, respectively. Integrating the latter over Minkowski spacetime leads to a topological winding number known as the second Chern number:
\begin{equation}
\frac{1}{8\pi^2}\int\mathrm{d}^4x\,\mathrm{tr}(G_{\mu\nu}\tilde{G}^{\mu\nu})=N\,,
\end{equation}
where $N\in\mathbb{Z}$.~'t Hooft showed that a gauge soliton, called the instanton, plays a crucial role in the nonconservation of chiral charge. The transition amplitude between different vacua encompassing such a charge violation is nonperturbative in the coupling constant~\cite{tHooft:1976rip}. A continuation of this analysis revealed that the following Lagrange density plays a role in the transition amplitude between two vacua of different topological charge~\cite{Callan:1976je}:
\begin{equation}
\label{eq:theta-term-SU(N)}
\mathcal{L}_{\theta}^{\mathrm{SU}(N)}=\frac{e^2\tilde{\theta}}{8\pi^2}\mathrm{tr}(G_{\mu\nu}\tilde{G}^{\mu\nu})\,,
\end{equation}
with the elementary charge $e$ and a phase $\tilde{\theta}\in[0,2\pi]$ characterizing a particular vacuum state. The Lagrange density is normalized such that the theory is invariant under $\tilde{\theta}\rightarrow \tilde{\theta}+2\pi$. For Maxwell electrodynamics, an analogous term can be proposed:
\begin{subequations}
\label{eq:lagrangian-theta}
\begin{align}
\mathcal{L}_{\theta}^{\mathrm{U(1)}}&=\frac{e^2\tilde{\theta}(x)}{8\pi^2}F_{\mu\nu}\tilde{F}^{\mu\nu}=:-\frac{\theta(x)}{2}F_{\mu\nu}\tilde{F}^{\mu\nu} \notag \\
&=2\theta(x)\mathbf{E}\cdot\mathbf{B}\,.
\end{align}
\end{subequations}
We promoted the phase $\tilde{\theta}$ of Eq.~\eqref{eq:theta-term-SU(N)} to a spacetime-dependent pseudoscalar field, which is odd under P and T: $\tilde{\theta}=\tilde{\theta}(x)$. For simplicity, a nontrivial factor has been absorbed into $\theta(x)$, which we will mostly be using in the remainder of the paper. After all, our current interest is in an effective description of material effects instead of developing suitable microscopic models. Note that Eq.~\eqref{eq:lagrangian-theta} is the coupling term of the axion to electromagnetic fields~\cite{Wilczek:1987mv}.

Indeed, Eq.~\eqref{eq:lagrangian-theta} follows from Eq.~\eqref{eq:lagrangian-modMax} by setting $(k_F)^{\mu\nu\varrho\sigma}=\theta(x)\varepsilon^{\mu\nu\varrho\sigma}$. Performing an integration by parts reveals the correspondence
\begin{subequations}
\label{eq:correspondence-CFJ-theta-theory}
\begin{align}
\mathcal{L}_{\mathrm{CFJ}}&=\mathcal{L}_{\theta}+\text{surface term}\,, \displaybreak[0] \\[1ex]
(k_{AF})_{\mu}&=\partial_{\mu}\theta\,, \displaybreak[0] \\[1ex]
\label{eq:correspondence-kAF}
(k_{AF})^0&=\dot{\theta}\,,\quad \mathbf{k}_{AF}=-\nabla\theta\,,
\end{align}
\end{subequations}
where we employed the homogeneous Maxwell equations, which remain unaffected by material properties. The emerging surface term vanishes in Minkowski spacetime.

The field equations of CFJ theory without external sources are
\begin{subequations}
\begin{equation}
\partial_{\alpha}F_{\mu}^{\phantom{\mu}\alpha}+(k_{AF})^{\alpha}\varepsilon_{\mu\alpha\beta\gamma}F^{\beta\gamma}=0\,,
\end{equation}
which can be recast into
\begin{equation}
\partial_{\alpha}F^{\alpha\mu}+2(k_{AF})_{\alpha}\tilde{F}^{\alpha\mu}=0\,.
\end{equation}
\end{subequations}
In terms of the physical fields, either of the latter two contains the following Maxwell equations:
\begin{subequations}
\label{eq:field-equations-CFJ}
\begin{align}
\nabla\cdot\mathbf{E}&=2\mathbf{k}_{AF}\cdot\mathbf{B}\,, \\[1ex]
-\frac{\partial\mathbf{E}}{\partial t}+\nabla\times\mathbf{B}&=2(k_{AF})^0\mathbf{B}-2\mathbf{k}_{AF}\times\mathbf{E}\,.
\end{align}
\end{subequations}
Properly chosen constitutive relations allow us to connect the CFJ field equations to the Maxwell equations in material media. They are proposed to be of the form
\begin{subequations}
\label{eq:constitutive-relations}
\begin{align}
\mathbf{D}&=\mathbf{E}+2\theta\mathbf{B}\,, \\[1ex]
\mathbf{H}&=\mathbf{B}-2\theta\mathbf{E}\,.
\end{align}
\end{subequations}
Inserting those into the inhomogeneous Maxwell equations without external sources provides
\begin{subequations}
\begin{equation}
0=\nabla\cdot\mathbf{D}=\nabla\cdot\mathbf{E}+2\nabla\theta\mathbf{B}\,,
\end{equation}
and
\begin{align}
0&=-\frac{\partial\mathbf{D}}{\partial t}+\nabla\times\mathbf{H} \notag \\
&=-\frac{\partial\mathbf{E}}{\partial t}+\nabla\times\mathbf{B}-2\dot{\theta}\mathbf{B}-2\theta\frac{\partial\mathbf{B}}{\partial t} \notag \\
&\phantom{{}={}}-2\nabla\theta\times\mathbf{E}-2\theta\nabla\times\mathbf{E} \notag \\
&=-\frac{\partial\mathbf{E}}{\partial t}+\nabla\times\mathbf{B}-2\dot{\theta}\mathbf{B}-2\nabla\theta\times\mathbf{E}\,.
\end{align}
\end{subequations}
These results correspond to Eq.~\eqref{eq:field-equations-CFJ}, when Eq.~\eqref{eq:correspondence-kAF} is taken into account. Comparing to the constitutive relations of Eq.~\eqref{eq:constitutive-relations-SME}, we identify
\begin{equation}
\label{eq:constitutive-properties-CFJ}
\epsilon=\mathds{1}_3=\mu\,,\quad \alpha=2\theta\mathds{1}_3\,.
\end{equation}
In contrast to $\kappa_{DB}$ and $\kappa_{HE}$ of Eq.~\eqref{eq:definition-kappaDB}, the latter $\alpha$ is not traceless. In fact, Colladay and Kosteleck\'{y} originally imposed the tracelessness on $\kappa_{DB}$ and $\kappa_{HE}$ via the Bianchi-type identity~\eqref{eq:bianchi-type-identity} of $k_F$ \cite{Colladay:1998fq}. The purpose was to avoid producing a $\theta$ term in modified Maxwell theory. Since for most SME treatments in Minkowski spacetime, background field components are taken as constants, the $\theta$ term is then a mere surface term that does not affect the field equations; cf.~Eq.~\eqref{eq:correspondence-CFJ-theta-theory}.

Optical media that have constitutive relations of the form of Eq.~\eqref{eq:constitutive-relations} are known as nonreciprocal, bi-isotropic media~\cite{Sihvola:1995}. For $\theta\in\mathbb{R}$, Tellegen media pose a prominent example of this class of materials. Interestingly, the name of the latter is a homage to Tellegen~\cite{Tellegen:1948}, who conceived an element of electric circuits beyond the conventional ones known at that time, which couples electric and magnetic fields with each other. Materials with $\mathrm{Im}(\theta)\neq 0$ satisfy the reciprocity condition of electrodynamics and are known as chiral media~\cite{Mun:2020}.

For $\theta\in\mathbb{R}$ to contribute to electromagnetic propagation, it must be an inhomogeneous function. Let us explore an interesting choice of $\theta$ that covers the features of certain Weyl semimetals~\cite{Zyuzin:2012tv,Armitage:2017cjs,Behrends:2018qkj,Kostelecky:2021bsb}. Their electron and hole dispersion relations for small wave vectors are Weyl cones separated in momentum space by a four-vector $b_{\mu}$. Here, $b_{\mu}$ denotes a pseudovector background field of the minimal-SME fermion sector~\cite{Colladay:1996iz,Colladay:1998fq,Kostelecky:2000mm}. There is a duality between the CFJ term and the $b_{\mu}$ term~\cite{Zyuzin:2012tv}, which leads to the intriguing finding~$\tilde{\theta}(x)=b_{\mu}x^{\mu}$. This contribution implies the chiral magnetic effect in Weyl semimetals~\cite{Burkov:2015hba,Silva:2020dli,Silva:2023ffk,Silva:2024mqn}. Note that Ref.~\cite{Silva:2023ffk} employs the crystallographic structure to infer properties of the magnetic-conductivity tensor. Their approach parallels the spirit of the analysis to be carried out below.

Since any material medium has a finite bulk, $\theta$ is at least position-dependent by definition. After all, $\theta$ can be constant in the bulk, but it will drop to zero outside the material. Then, nontrivial boundary contributions are expected to occur. For example, the choice $\tilde{\theta}(x)=\pi[1-\Theta(z)]$ with the Heaviside function $\Theta(z)$ governs the bulk of a topological insulator for $z<0$ when T symmetry is preserved~\cite{Sekine:2020ixs}. 

In fact, the Lagrange density of the $\theta$ term in Eq.~\eqref{eq:lagrangian-theta} can be generalized to be anisotropic. According to Eq.~\eqref{eq:lagrange-density-physical-fields},
\begin{equation}
\mathcal{L}_{\mathrm{modMax}}\supset \mathbf{E}\cdot \kappa_{DB}\cdot \mathbf{B}\,,
\end{equation}
i.e., the SME coefficients that are noninvariant under P and T simultaneously couple the electric and magnetic fields with each other, as expected. By including the $\theta$ term,
\begin{equation}
\label{eq:magnetoelectric-coupling-generic}
\mathcal{L}_{\theta}^{\mathrm{U}(1)}+\mathcal{L}_{\mathrm{modMax}}\supset \left[2\theta(x)\delta^{ij}+(\kappa_{DB})^{ij}\right]E^iB^j\,,
\end{equation}
where $\kappa_{DB}$ is the sum of Eqs.~\eqref{eq:kappa-DB-sector-1} and \eqref{eq:kappa-DB-sector-2}. Recall that $\kappa_{DB}$ is $\mathfrak{so}(3)$-valued for the first principal sector. For the second principal sector, $\kappa_{DB}$ is decomposed into an antisymmetric and a symmetric piece. Then,
\begin{subequations}
\begin{align}
\mathcal{L}_{\theta}^{\mathrm{U}(1)}+\mathcal{L}_{\mathrm{modMax}}&\supset \left[2\,\mathbf{E}\cdot g\cdot\mathbf{B}+(\mathbf{E}\times\mathbf{B})\cdot(\tilde{\mathbf{k}}-\tilde{\boldsymbol{\kappa}})\right]\,, \displaybreak[0]\\[1ex]
g&=\begin{pmatrix}
\theta+k^2 & -k^9/2 & k^8/2 \\
-k^9/2 & \theta-k^1 & k^{10}/2 \\
k^8/2 & k^{10}/2 & \theta+k^1-k^2 \\
\end{pmatrix}\,, \displaybreak[0]\\[1ex]
\tilde{\boldsymbol{\kappa}}&=\begin{pmatrix}
\tilde{\kappa}^{01} \\
\tilde{\kappa}^{02} \\
\tilde{\kappa}^{03} \\
\end{pmatrix}\,,\quad \tilde{\mathbf{k}}=\begin{pmatrix}
-k^{10} \\
k^8 \\
k^9 \\
\end{pmatrix}\,,
\end{align}
\end{subequations}
with a skewed three-dimensional metric $g^{ij}$ and two spatial vectors $\tilde{\boldsymbol{\kappa}}$ and $\tilde{\mathbf{k}}$. Thus, the SME provides natural generalizations of the $\theta$ term of Eq.~\eqref{eq:lagrangian-theta}. While the $\theta$ term describes an isotropic magnetoelectric coupling, the coefficients contained in $\kappa_{DB}$ can parametrize possible anisotropies. This gives rise to materials known as isotropic-anisotropic or bi-anisotropic, depending on the properties of permittivity and permeability. Electromagnetic propagation in all these classes of optical media has been subject to intensive studies; see, e.g., Refs.~\cite{Silva:2022sps,Silva:2022bnv}.

\subsection{Decompositions}

The previous results and discussions lead to the following decomposition of coefficients in the minimal-SME electromagnetic sector:
\begin{align}
\underbrace{\text{SME coefficients}}_{21}&=\underbrace{\text{1${}^{\text{st}}$ sector}}_{\tilde{\kappa}^{\mu\nu}\,(9)}+\underbrace{\text{2${}^{\text{nd}}$ sector}}_{k^a\,(10)} \notag \\
&\phantom{{}={}}+\underbrace{\text{Double trace}}_{k_{\mathrm{tr}}\,(1)}+\underbrace{\text{Pseudoscalar}}_{\theta\,(1)}\,.
\end{align}
Let us compare the latter directly to the material parameters of electrodynamics:
\begin{align}
\underbrace{\begin{array}{c}
\text{Material} \\
\text{parameters}
\end{array}}_{21}&=\underbrace{\text{Permittivity}}_{\epsilon_{\{ij\}}\,(6)}+\underbrace{\text{Permeability}}_{\mu_{\{ij\}}\,(6)} \notag \\
&\phantom{{}={}}+\underbrace{\begin{array}{c}
\text{Magnetoelectric} \\
\text{couplings}
\end{array}}_{\alpha_{ij}\,(9)}\,,
\end{align}
where a pair of curly brackets indicates symmetrization over the enclosed indices. The permittivity and permeability tensors are symmetric, by definition, in contrast to the tensor of magnetoelectric couplings. The number of SME coefficients matches the number of parameters, as expected. This again emphasizes the significance of $k_{\mathrm{tr}}$ when dealing with material effects.

\section{Crystallographic and magnetic point groups}
\label{eq:crystallographics-groups}

The CPT-even part of the electromagnetic sector in the SME involves certain configurations that preserve PT symmetry while violating P and T symmetries on their own. This partial conservation of certain discrete symmetries has observable consequences in electromagnetic-wave propagation, such as anisotropic modifications of the lightcone in the medium. The latter become manifest as direction-dependent propagation effects as well as birefringence. Crystal structures that violate either P or T symmetry exhibit different features determined by the broken discrete symmetries. As mentioned above, $\kappa_{DB}$ violates P and T individually but preserves PT. A possible connection is drawn to magnetoelectric crystals where the simultaneous breaking of P and T enables cross-couplings between electric and magnetic fields \cite{Fiebig:2005}.

Chiral and parity symmetries are profoundly related to each other, as revealed by the optical activity of crystals~\cite{Barron:2009}. Due to the properties of spatial inversion, centrosymmetric crystals cannot support intrinsic chirality, which excludes optical activity. Under a parity transformation,
\begin{equation}
\mathcal{P}: \psi_L(\mathbf{r}) \rightarrow \psi_R(-\mathbf{r}) \neq \psi_L(\mathbf{r})\,,
\end{equation}
where $\psi_L$ and $\psi_R$ represent left- and right-handed chiral states, respectively. For optical activity to exist, the refractive indices of the material associated with different chiralities must differ from each other:
\begin{equation}
n_L(\omega) \neq n_R(\omega)\,.
\end{equation}
This condition becomes impossible to satisfy when P is preserved, which requires that $n_L=n_R$.

The microscopic charge and current densities $\rho_m(\mathbf{r})$ and $\mathbf{j}_m(\mathbf{r})$, respectively, are fundamental in determining the electromagnetic properties of a material. Under T, the current reverses its direction, $\mathbf{j}_m\rightarrow -\mathbf{j}_m$. Thereupon, a time-reversal invariant crystal requires a configuration with $\mathbf{j}_m(\mathbf{r})\equiv 0$ at all $\mathbf{r}$. Then, both the time-averaged magnetic field and magnetic moment vanish throughout the crystal lattice. Hence, such crystals are inherently nonmagnetic, which is something observed in a vast number of crystalline materials \cite{Landau:1984}.

If we want to describe electromagnetic properties of crystals comprehensively, we must go beyond basic P and T operations. Crystallography relies on crystallographic point groups for nonmagnetic materials and magnetic point groups for systems with magnetic ordering. They tell us how the symmetries present in both the crystal structure and spin configuration are linked to observable physical phenomena. The symmetry groups characterize the transformations that leave invariant both the crystal lattice and the organization of magnetic moments. Therefore, they are capable of associating properties described by $k_F$ with each class of crystals. Performing a systematic analysis based on group-theory methods reveals how symmetry operations in absence or combination of T govern phenomena on macroscopic scales such as optical activity, magnetoelectric effects, and birefringence.

\subsection{Covariant formulation of symmetry transformations}

Analyzing the symmetry properties of $k_F$ requires that the operations of magnetic point groups be expressed covariantly. A relativistic symmetry operation is represented by a linear transformation in Minkowski spacetime, i.e., a matrix $\Lambda^{\mu}_{\phantom{\mu}\nu}$ that preserves the relativistic scalar product $x\cdot y=x^{\mu}y_{\mu}$. Accordingly, for the fourth-rank tensor $k_F$, the covariant transformation takes the general form
\begin{equation}
(k_F)'_{\mu\nu\rho\sigma} =
\Lambda_{\mu}^{\phantom{\mu}\alpha}\,
\Lambda_{\nu}^{\phantom{\nu}\beta}\,
\Lambda_{\rho}^{\phantom{\rho}\gamma}\,
\Lambda_{\sigma}^{\phantom{\sigma}\delta}
(k_F)_{\alpha\beta\gamma\delta}\,.
\label{eq:kF-transform}
\end{equation}
The tensor is invariant under a given transformation if
\begin{equation}
(k_F)'_{\mu\nu\rho\sigma} = (k_F)_{\mu\nu\rho\sigma}\,,
\label{leigeral}
\end{equation}
which imposes constraints on the controlling coefficients $(k_F)_{\mu\nu\rho\sigma}$ and determines the type of emergent Lorentz breaking compatible with the symmetry group considered.

Note that for improper spatial transformations such as reflections or inversions, as well as for time reversal, we must distinguish between polar and axial tensors (pseudotensors). Considering a linear spatial transformation governed by, say, $\mathcal{D}_{ij}$ with spatial indices $i$ and $j$, a pseudotensor acquires an additional factor of $\det(\mathcal{D})$ and is also odd under T. In the electromagnetic sector, in particular, we have that
\begin{subequations}
\begin{align}
\mathcal{P}:\ \mathbf{E}&\!\to\!-\mathbf{E}\,,\quad \mathbf{B}\!\to\!+\mathbf{B}\,, \\[1ex]
\mathcal{T}:\ \mathbf{E}&\!\to\!+\mathbf{E}\,,\quad \mathbf{B}\!\to\!-\mathbf{B}\,,
\end{align}
\end{subequations}
i.e., $\mathbf{B}$ is even under P and odd under T, whereas $\mathbf{E}$ is odd under P and even under T.

We are now ready to analyze how the four matrices of Eq.~\eqref{eq:constitutive-relations-fields} transform under the symmetry operations of the magnetic point groups. First, we will examine their behavior under spatial symmetry transformations that include both proper and improper operations such as rotations and reflections, respectively. Time reversal will be taken into account, too, where particular attention must be paid to distinct transformation behaviors of polar and axial components. This approach allows us to derive covariant transformation laws for $k_F$, which are the foundation to grant consistency with the symmetries of the atomic lattice. The transformations considered can be expressed in covariant form as
\begin{equation}
(\Lambda^{\mu}_{\phantom{\mu}\nu})=\begin{pmatrix}
1 & 0 \\
0 & (\Lambda_{ij}) \\
\end{pmatrix}\,,
\end{equation}
with the purely spatial transformation $(\Lambda_{ij})$. Here, it is understood that the timelike and spacelike components of a four-vector do not mix with each other. After all, boosts are not important for the setting to be developed.

For the constitutive tensor defined in Eq.~\eqref{eq:definition-kappaDB} a spatial transformation leads to
\begin{equation}
(\kappa_{DB}')^{lm}=(k_{F}')^{0lrs}\epsilon'^{mrs}\,,
\end{equation}
where the Levi-Civita tensor transforms as
\begin{equation}
\epsilon'^{mrs}=\det(\Lambda) \Lambda^{m}_{\phantom{m}k} \Lambda^{r}_{\phantom{r}p} \Lambda^{s}_{\phantom{s}q} \epsilon^{kpq}\,.
\label{LeviCivita}
\end{equation}
Thus, using Eq.~\eqref{LeviCivita}, the transformation law reads
\begin{equation}
(\kappa_{DB}')^{lm}=\det(\Lambda) \Lambda^{l}_{\phantom{l}j} \Lambda^{m}_{\phantom{m}k} (\kappa_{DB})^{jk}\,.
\end{equation}
The other constitutive tensors of Eqs.~\eqref{eq:definition-kappaDE} and \eqref{eq:definition-kappaHB} are polar quantities, whereupon their transformation laws do not involve the determinant of the transformation matrix:
\begin{subequations}
\begin{align}
(\kappa_{DE}')^{lm} & =\Lambda^{l}_{\phantom{l}j}\Lambda^{m}_{\phantom{m}k}(\kappa_{DE})^{jk}\,, \\
(\kappa_{HB}')^{lm} & =\Lambda^{l}_{\phantom{l}j}\Lambda^{m}_{\phantom{m}k}(\kappa_{HB})^{jk}\,.
\end{align}
\end{subequations}
Note that the difference between the transformation law of polar and axial tensors becomes evident for improper transformations, i.e., reflections, spatial inversion, and time reversal. Contrary to $\kappa_{DE}$ and $\kappa_{HB}$, the tensor $\kappa_{DB}$ then acquires an additional minus sign. Thus, the latter changes sign under both T and P.

\subsection{Electric and magnetic susceptibilities}

\begin{table*}[t]
\centering
\begin{tabular}{ccm{5cm}ccc} 
    \toprule
    Set & Crystal & \multicolumn{1}{c}{Classes} & \# Coeffs. & Allowed form of $\kappa_{DE}$ \\
    \midrule
    I & Triclinic & \centering \textbf{1}, $\bar{\mathbf{1}}$ & $6+1$ & $\begin{pmatrix}
        \tilde{\kappa}^{00}-\tilde{\kappa}^{11} & -\tilde{\kappa}^{12} & -\tilde{\kappa}^{13} \\
        -\tilde{\kappa}^{12} & \tilde{\kappa}^{00}-\tilde{\kappa}^{22} & -\tilde{\kappa}^{23} \\
        -\tilde{\kappa}^{13} & -\tilde{\kappa}^{23} & \tilde{\kappa}^{11}+\tilde{\kappa}^{22}
    \end{pmatrix}+\frac{k_{\mathrm{tr}}}{6}\mathds{1}_3$ \\[0.5cm]
    II & Monoclinic & \centering \textbf{2}, \textbf{m}, \textbf{2}/\textbf{m} & $4+1$ &
    $\begin{pmatrix}
        \tilde{\kappa}^{00}-\tilde{\kappa}^{11} & 0 & -\tilde{\kappa}^{13} \\
        0 & \tilde{\kappa}^{00}-\tilde{\kappa}^{22} & 0 \\
        -\tilde{\kappa}^{13} & 0 & \tilde{\kappa}^{11}+\tilde{\kappa}^{22}
    \end{pmatrix}+\frac{k_{\mathrm{tr}}}{6}\mathds{1}_3$ \\
    III & Orthorhombic & \centering \textbf{222}, \textbf{mm2}, \textbf{mmm} & $3+1$ & $\mathrm{diag}\Big(\tilde{\kappa}^{00}-\tilde{\kappa}^{11},\tilde{\kappa}^{00}-\tilde{\kappa}^{22},\tilde{\kappa}^{11}+\tilde{\kappa}^{22}\Big)+\frac{k_{\mathrm{tr}}}{6}\mathds{1}_3$ \\
    IV & Uniaxial &
    \centering \begin{minipage}{4cm}
    \centering \textbf{3}, $\bar{\mathbf{3}}$, \textbf{32}, \textbf{3m}, $\bar{\mathbf{3}}$\textbf{m},
    \textbf{4}, $\bar{\mathbf{4}}$, \textbf{4}/\textbf{m}, \textbf{422}, \textbf{4mm},
    $\bar{\mathbf{4}}$\textbf{2m}, \textbf{4}/\textbf{mmm}, \textbf{6}, $\bar{\mathbf{6}}$, \textbf{6}/\textbf{m},
    \textbf{622}, \textbf{6mm}, $\bar{\mathbf{6}}$\textbf{m2}, \textbf{6}/\textbf{mmm} 
    \end{minipage} & $2+1$ &
    $\mathrm{diag}\Big(\tilde{\kappa}^{00}-\tilde{\kappa}^{11},\tilde{\kappa}^{00}-\tilde{\kappa}^{11},2\tilde{\kappa}^{11}\Big)+\frac{k_{\mathrm{tr}}}{6}\mathds{1}_3$ \\[5ex]
    V & Cubic & \centering \textbf{23}, \textbf{m3}, \textbf{432}, $\bar{\mathbf{4}}$\textbf{3m}, \textbf{m}$\bar{\mathbf{3}}$\textbf{m} & $1+1$ &
    $\left(\tilde{\kappa}_{\mathrm{tr}}+\frac{k_{\mathrm{tr}}}{6}\right)\mathds{1}_3$ \\
    \bottomrule
\end{tabular}
\caption{SME electric-susceptibility tensor $\kappa_{DE}$ of Eq.~\eqref{eq:definition-kappaDE} consistent with crystallographic point groups. The first column numerates the sets of the classification. The second column provides information on the type of crystal lattice, and the third lists the symmetry class. The fourth column gives the number of independent nonzero SME coefficients, and the last one states the forms of $\kappa_{DE}$ that the symmetry classes dictate.}
\label{tab:electric_susceptibility}
\end{table*}
\begin{table*}[t]
\centering
\begin{tabular}{ccm{4cm}ccc}
    \toprule
     Set & Crystal & \centering Classes & \# Coeffs. & Allowed form of $\kappa_{HB}$ \\
    \midrule
    I & Triclinic & \centering \textbf{1}, $\bar{\mathbf{1}}$ & $6+5+1$ &
    $-\kappa_{DE}+\frac{k_{\mathrm{tr}}}{3}\mathds{1}_3-2\begin{pmatrix}
        -(k^{3}+k^{4}) & k^{5} & k^{6} \\
        k^{5} & k^{3} & k^{7} \\
        k^{6} & k^{7} & k^{4} \\
    \end{pmatrix}$ \\[0.5cm]
    II & Monoclinic & \centering \textbf{2}, \textbf{m}, \textbf{2}/\textbf{m} & $4+3+1$ &
    $-\kappa_{DE}+\frac{k_{\mathrm{tr}}}{3}\mathds{1}_3-2\begin{pmatrix}
       -(k^{3}+k^{4}) & 0 & k^{6} \\
        0 & k^{3} & 0 \\
        k^{6} & 0 & k^{4} \\
    \end{pmatrix}$ \\[0.5cm]
    III & Orthorhombic & \centering \textbf{222}, \textbf{mm2}, \textbf{mmm}
    & $3+2+1$ &
    $-\kappa_{DE}+\frac{k_{\mathrm{tr}}}{3}\mathds{1}_3-2\,\mathrm{diag}\Big(-(k^{3}+k^{4}),k^{3},k^{4}\Big)$ \\
    IV & Uniaxial &
    \centering \begin{minipage}{4cm}
    \centering \textbf{3}, $\bar{\mathbf{3}}$, \textbf{32}, \textbf{3m},
    $\bar{\mathbf{3}}$\textbf{m}, \textbf{4}, $\bar{\mathbf{4}}$, \textbf{4}/\textbf{m},
    \textbf{422}, \textbf{4mm}, $\bar{\mathbf{4}}$\textbf{2m},
    \textbf{4}/\textbf{mmm}, \textbf{6}, $\bar{\mathbf{6}}$,
    \textbf{6}/\textbf{m}, \textbf{622}, \textbf{6mm},
    $\bar{\mathbf{6}}$\textbf{m2}, \textbf{6}/\textbf{mmm}
    \end{minipage} & $2+1+1$ &
    $-\kappa_{DE}+\frac{k_{\mathrm{tr}}}{3}\mathds{1}_3-2\,\mathrm{diag}\Big(k^{3},k^{3},-2k^{3}\Big)$ \\[5ex]
    V & Cubic & \centering \textbf{23}, \textbf{m3}, \textbf{432},
    $\bar{\mathbf{4}}$\textbf{3m}, \textbf{m}$\bar{\mathbf{3}}$\textbf{m} & $1+1$ &
    $\left(-\tilde{\kappa}_{\mathrm{tr}}+\frac{k_{\mathrm{tr}}}{6}\right)\mathds{1}_3$ \\
    \bottomrule
\end{tabular}
\caption{The same as Tab.~\ref{tab:electric_susceptibility}, but for the SME magnetic-susceptibility tensor $\kappa_{HB}$ of Eq.~\eqref{eq:definition-kappaHB}. Each $\kappa_{HB}$ is expressed in terms of the appropriate $\kappa_{DE}$ that Tab.~\ref{tab:electric_susceptibility} provides. In contrast to $\kappa_{DE}$, the matrices $\kappa_{HB}$ also depend on the coefficients of the second principal sector of $k_F$. Therefore, the fourth column states the numbers of independent nonzero SME coefficients of the first and second principal sectors.}
\label{tab:magnetic_susceptibility}
\end{table*}

The electric and magnetic susceptibilities $\chi^P$ and $\chi^M$, respectively, as well as the SME quantities $\kappa_{DE}$ and $\kappa_{HB}$ are both polar and invariant under time reversal. Therefore, the 32 crystallographic point groups entirely determine their possible matrix representations. First of all, let us study $\kappa_{DE}$. Note that coefficients of the second principal sector do not contribute, whereupon the full $\kappa_{DE}$ is already given in Eq.~\eqref{eq:kappa-DE-sector1}.

In the following, the generators for point groups are indispensable. These can be found, e.g., in Tab.~4.2 of Ref.~\cite{Newnham:2005}, which is used in conjunction with Tab.~4.1.
Let $\mathcal{R}^{[hkl]}(\Omega)$ describe a rotation with angle $\Omega$ around the crystallographic direction~$[hkl]$; cf.~Eq.~\eqref{eq:generic-rotation} for a possible representation. Moreover, $\mathcal{M}^{[hkl]}$ denotes a reflection at a mirror perpendicular to~$[hkl]$. Matrix elements of the transformation matrices are labeled by Latin letters $i,j,\dots$ and refer to the crystalline basis.

For demonstration purpose, we consider two examples: the groups \textbf{2}/\textbf{m} and \textbf{6}/\textbf{m}, respectively. The generators of the former are 2-fold rotations around $[010]$ and reflections at a mirror perpendicular to this very same axis:
\begin{subequations}
\begin{align}
\mathcal{R}^{[010]}(\pi)&=\mathrm{diag}(-1,1,-1)\,, \\[1ex]
\mathcal{M}^{[010]}&=\mathrm{diag}(1,-1,1)\,.
\end{align}
\end{subequations}
Then, the following transformations must be analyzed:
\begin{subequations}
\begin{align}
(\kappa_{DE}')^{ij}&=\mathcal{R}^{[010]}_{il}(\pi)(\kappa_{DE})^{lm}\mathcal{R}^{[010]}_{jm}(\pi)\,, \\[1ex]
(\kappa_{DE}'')^{ij}&=\mathcal{M}^{[010]}_{il}(\kappa_{DE})^{lm}\mathcal{M}^{[010]}_{jm}\,.
\end{align}
\end{subequations}
For invariance, we demand that $\kappa_{DE}'=\kappa_{DE}$ as well as $\kappa_{DE}''=\kappa_{DE}$. Each requirement eliminates the same coefficients $\tilde{\kappa}^{12}$ and $\tilde{\kappa}^{23}$, which are incompatible. Therefore, the result is $\kappa_{DE}$ of the second line of Tab.~\ref{tab:electric_susceptibility}.

As a second example, consider the high-symmetry group \textbf{6}/\textbf{m}, which involves a 6-fold rotation around the third axis and reflections at a mirror orthogonal to this axis:
\begin{subequations}
\begin{align}
\mathcal{R}^{[001]}\left(\frac{\pi}{3}\right)&=\begin{pmatrix}
1/2 & \sqrt{3}/2 & 0 \\
-\sqrt{3}/2 & 1/2 & 0 \\
0 & 0 & 1 \\
\end{pmatrix}\,, \\[1ex]
\mathcal{M}^{[001]}&=\mathrm{diag}(1,1,-1)\,.
\end{align}
\end{subequations}
In an analogous manner, we must look into the transformations
\begin{subequations}
\begin{align}
\label{eq:requirement-6m-1}
(\kappa_{DE}')^{ij}&=\mathcal{R}^{[001]}_{il}\left(\frac{\pi}{3}\right)(\kappa_{DE})^{lm}\mathcal{R}^{[001]}_{jm}\left(\frac{\pi}{3}\right)\,, \\[1ex]
\label{eq:requirement-6m-2}
(\kappa_{DE}'')^{ij}&=\mathcal{M}^{[001]}_{il}(\kappa_{DE})^{lm}\mathcal{M}^{[001]}_{jm}\,.
\end{align}
\end{subequations}
In fact, Eq.~\eqref{eq:requirement-6m-1} under the requirement $\kappa_{DE}'=\kappa_{DE}$ eliminates $\tilde{\kappa}^{12}$, $\tilde{\kappa}^{13}$, and $\tilde{\kappa}^{23}$. It also enforces that $\tilde{\kappa}^{11}=\tilde{\kappa}^{22}$. Interestingly, $\kappa_{DE}''=\kappa_{DE}$ based on Eq.~\eqref{eq:requirement-6m-2} is automatically satisfied when the previous restrictions are implemented. This leads to $\kappa_{DE}$ in the fourth line of Tab.~\ref{tab:electric_susceptibility}.
\begin{table*}
\centering
\begin{tabular}{cccccccccccc}
\toprule
Class & Mineral & \multicolumn{9}{c}{Nonzero SME coefficients} & Entry in database \\
\midrule
      &         & $\tilde{\kappa}^{00}$ & $\tilde{\kappa}^{11}$ & $\tilde{\kappa}^{22}$ & $k^1$ & $k^2$ & $k^3$ & $k^4$ & $k_{\mathrm{tr}}$ & $\theta$ \\
\cmidrule(lr){3-11}
\textbf{2}/$\mathbf{m}$ & Datolite & $1.29$ & $0.51$ & $0.42$ & 0 & 0 & $-0.01$ & $-0.03$ & $4\tilde{\kappa}^{00}$ & 0 & 1340 \\
\textbf{222} & Edingtonite & $1.04$ & $0.38$ & $0.34$ & 0 & 0 & $0.00$ & $-0.01$ & $4\tilde{\kappa}^{00}$ & 0 & 1353 \\
\textbf{mm2} & Hemimorphite & $1.22$ & $0.44$ & $0.43$ & 0 & 0 & $0.01$ & $-0.02$ & $4\tilde{\kappa}^{00}$ & 0 & 1860 \\
\textbf{mmm} & Andalusite & $1.26$ & $0.44$ & $0.42$ & 0 & 0 & $0.00$ & $-0.01$ & $4\tilde{\kappa}^{00}$ & 0 & 217 \\
\textbf{32} & Cinnabar & $6.12$ & $2.76$ & $\tilde{\kappa}^{11}$ & 0 & 0 & $0.36$ & $-2k^3$ & $4\tilde{\kappa}^{00}$ & 0 & 1052 \\
$\bar{\mathbf{4}}$\textbf{3m} & Sphalerite & $3.46$ & $\tilde{\kappa}^{00}/3$ & $\tilde{\kappa}^{00}/3$ & 0 & 0 & 0 & 0 & $4\tilde{\kappa}^{00}$ & 0 & 3727 \\
\textbf{m}$\bar{\mathbf{3}}$\textbf{m} & Cuprite & $5.34$ & $\tilde{\kappa}^{00}/3$ & $\tilde{\kappa}^{00}/3$ & 0 & 0 & 0 & 0 & $4\tilde{\kappa}^{00}$ & 0 & 1172 \\
\midrule
$\bar{\mathbf{3}}$'$\mathbf{m}$' & Eskolaite & $8.88$ & $2.49$ & $\tilde{\kappa}^{11}$ & $0.57$ & $-k^1$ & $-0.23$ & $-2k^3$ & $4\tilde{\kappa}^{00}$ & $0.93$ & 1411 \\
\bottomrule
\end{tabular}
\caption{Examples for crystals whose electric permittivity and permeability tensors $\epsilon$ and $\mu$, respectively, of Eq.~\eqref{eq:permittivity} are expressed in terms of SME coefficients. The first column lists the symmetry class and the second the mineral. The third column provides the nonzero SME coefficients, where $\tilde{\kappa}^{33}$ is fixed by the tracelessness of $\tilde{\kappa}$. Finally, references to the database \href{http://www.mindat.org}{www.mindat.org} are stated where experimental data on refractive indices can be found, which imply the proper eigenvalues of $\epsilon$. Since the materials are nonmagnetic, $\mu=\mathds{1}_3$. Note that the material in the last line has magnetoelectric properties.}
\label{tab:example-materials}
\end{table*}

Applying the aforementioned procedure to the remaining 30 crystallographic groups, implies compatible forms of $\kappa_{DE}$ provided in Tab.~\ref{tab:electric_susceptibility}. Several comments are in order. First, there are only five different forms for $\kappa_{DE}$, i.e., a given form of the latter matrix can be compatible with different symmetries. Second, the number of independent SME coefficients reduces with increasing crystal symmetry from the maximum of 7 to 2 coefficients remaining for the isotropic case. Note that there is no $\kappa_{DE}$ with 6 independent coefficients. Third, the matrices for the first two sets of point groups are nondiagonal. However, since these matrices are symmetric, they can be diagonalized by finding the principal axes. As of the third set of point groups, $\kappa_{DE}$ is automatically diagonal. Fourth, cubic crystals, which exhibit isotropy, are characterized by the isotropic coefficient \cite{Kostelecky:2002hh}
\begin{equation}
\label{eq:definition-kappatr}
\tilde{\kappa}_{\mathrm{tr}}:=\frac{1}{3}(\kappa_{DE})^{ll}=\frac{2}{3}\tilde{\kappa}^{00}\,,
\end{equation}
as well as the double trace $k_{\mathrm{tr}}$, as expected. Note that $\tilde{\kappa}_{\mathrm{tr}}$ is not to be confused with the double trace $k_{\mathrm{tr}}$, which is a scalar under the emergent Lorentz group and was introduced at the end of Sec.~\ref{sec:principal-sectors}.

An analogous procedure applies to $\kappa_{HB}$. In contrast to $\kappa_{DE}$ studied previously, $\kappa_{HB}$ depends on coefficients from both principal sectors of $k_F$. Its complete matrix representation follows from Eqs.~\eqref{eq:kappa-HB-sector1}, \eqref{eq:kappa-HB-sector2} and is given by
\begin{equation}
\kappa_{HB}=-\kappa_{DE}+\frac{k_{\mathrm{tr}}}{3}\mathds{1}_3-2\left(
\begin{array}
[c]{ccc}%
-(k^3+k^4) & k^{5} & k^{6}\\
k^{5} & k^{3} & k^{7}\\
k^{6} & k^{7} & k^{4} \\
\end{array}
\right)\,,
\end{equation}
with $\kappa_{DE}$ stated in the first line of Tab.~\ref{tab:electric_susceptibility}. This relationship is a consequence of how $k_F$ is decomposed into its two principal sectors.
The previous results for $\kappa_{DE}$ still hold, which is why the analysis can be restricted to the matrix of coefficients of the second principal sector. The outcome is to be found in Tab.~\ref{tab:magnetic_susceptibility}. Note that for cubic crystal lattices, $\tilde{\kappa}_{\mathrm{tr}}$ enters $\kappa_{DE}$ and $\kappa_{HB}$ with opposite signs, whereas the signs are the same for $k_{\mathrm{tr}}$. This property is an implication of $\tilde{\kappa}_{\mathrm{tr}}$ changing the slope of the light cone, as compared to the vacuum, whereas $k_{\mathrm{tr}}$ preserves Lorentz invariance.

The components of the second principal sector are critical, as they generate the degrees of freedom necessary to distinguish electric from magnetic susceptibility. This distinction must be made because many crystals are nonmagnetic, which renders the magnetic permeability trivial. Thus, $\kappa_{HB}$ accounts for a magnetic response that differs fundamentally from the electric one, even when this response is weak or isotropic. The behavior of $\kappa_{HB}$ for increasing symmetry of the crystal lattice is similar to that of $\kappa_{DE}$.

\subsubsection{Application to real minerals}

Tables~\ref{tab:electric_susceptibility} and \ref{tab:magnetic_susceptibility} are valuable to parametrize electromagnetic properties of real minerals found in nature. To do so, the point group of the crystal lattice must be identified. Depending on the point group, the matrices $\epsilon$ and $\mu$ are nondiagonal, whereupon they should be diagonalized. The reason is that experimental values such as refractive indices are always given in the system of principal axes.

It is possible to diagonalize the full forms of $\epsilon$ and $\mu$ for the simplest groups \textbf{1} and $\bar{\mathbf{1}}$. However, these eigenvalues are intricate, which makes any consideration challenging. Diagonalization for the groups of set II is feasible, and $\epsilon,\mu$ are automatically diagonal for the remaining sets.

Once $\epsilon$ and $\mu$ are given in their diagonal forms, the square roots of the eigenvalues of $\epsilon$ correspond to the refractive indices of the crystal according to the generic relationship $n=\sqrt{\epsilon}$. The majority of crystals are nonmagnetic, whereupon we take into account that $\mu=\mathds{1}_3$. The resulting system of coupled algebraic equations must be solved for the SME coefficients. The equations can be nonlinear, but solutions have been obtained for each specific case studied. Let us discuss their properties as follows.

Our examples start with the point symmetry groups of set II, which should be suitable to describe materials with a generic $\epsilon=\mathrm{diag}(\epsilon_1,\epsilon_2,\epsilon_3)$. Note that $\kappa_{DE}$ and $\kappa_{HB}$ of Tabs.~\ref{tab:electric_susceptibility} and \ref{tab:magnetic_susceptibility} are both nondiagonal in this case. Let $\varphi_{ij}$ be the angle between the axes $i$ and $j$. For the unit cell of the monoclinic crystal lattice, it holds that $\varphi_{xy}=\varphi_{yz}=90^{\circ}$, whereas $\varphi_{xz}>90^{\circ}$. Permittivity and permeability are diagonal in the system of principal axes, and the values of refractive indices in the literature are conventionally stated in this system. Therefore, we must perform a rotation by an angle $\Omega=\varphi_{xz}-90^{\circ}$ back to the Cartesian coordinate system, where $\epsilon$ is nondiagonal and has the form of the latter $\kappa_{DE}$ and $\kappa_{HB}$. Doing so, leads to
\begin{equation}
\epsilon'=\begin{pmatrix}
\epsilon_1\cos^2\Omega+\epsilon_3\sin^2\Omega & 0 & \frac{1}{2}(\epsilon_3-\epsilon_1)\sin(2\Omega) \\
0 & \epsilon_2 & 0 \\
\frac{1}{2}(\epsilon_3-\epsilon_1)\sin(2\Omega) & 0 & \epsilon_1\sin^2\Omega+\epsilon_3\cos^2\Omega \\
\end{pmatrix}\,.
\end{equation}
Now, solving the system of equations $\epsilon'=\mathds{1}_3+\kappa_{DE}$ and $\mu=\mathds{1}_3+\kappa_{HB}$ for the SME coefficients implies
\begin{subequations}
\label{eq:solution-group-2}
\begin{align}
\tilde{\kappa}^{00}&=\frac{1}{4}(\mathrm{tr}(\epsilon)-3)\,,\quad k_{\mathrm{tr}}=4\tilde{\kappa}^{00}\,, \displaybreak[0]\\[2ex]
\tilde{\kappa}^{11}&=\frac{1}{12}\Big\{-\mathrm{tr}(\epsilon)+6[(\epsilon_3-\epsilon_1)\cos(2\Omega)+\epsilon_2]\Big\}-\frac{1}{4}\,, \displaybreak[0]\\[2ex]
\tilde{\kappa}^{22}&=\frac{5}{12}\mathrm{tr}(\epsilon)-\epsilon_2-\frac{1}{4}\,, \\[2ex]
\tilde{\kappa}^{13}&=\frac{1}{2}(\epsilon_1-\epsilon_3)\sin(2\Omega)\,, \displaybreak[0]\\[2ex]
k^3&=\frac{1}{6}\mathrm{tr}(\epsilon)-\frac{\epsilon_2}{2}\,, \displaybreak[0]\\[2ex]
k^4&=\frac{1}{12}\Big\{-\mathrm{tr}(\epsilon)+3[(\epsilon_1-\epsilon_3)\cos(2\Omega)+\epsilon_2]\Big\}\,, \displaybreak[0]\\[2ex]
k^6&=\frac{1}{4}(\epsilon_1-\epsilon_3)\sin(2\Omega)\,.
\end{align}
\end{subequations}
For the point groups of set III, the matrices $\kappa_{DE}$ and $\kappa_{HB}$ are automatically diagonal. Hence, we can work in the system or principal axes. The solution then takes the following form:
\begin{subequations}
\begin{align}
\tilde{\kappa}^{00}&=\frac{1}{4}(\mathrm{tr}(\epsilon)-3)\,,\quad k_{\mathrm{tr}}=4\tilde{\kappa}^{00}\,, \\[2ex]
\tilde{\kappa}^{11}&=\frac{5}{12}\mathrm{tr}(\epsilon)-\epsilon_1-\frac{1}{4}\,,\quad \tilde{\kappa}^{22}=\frac{5}{12}\mathrm{tr}(\epsilon)-\epsilon_2-\frac{1}{4}\,, \displaybreak[0]\\[1ex]
k^{3}&=\frac{1}{6}\mathrm{tr}(\epsilon)-\frac{\epsilon_2}{2}\,,\quad k^4=\frac{1}{6}\mathrm{tr}(\epsilon)-\frac{\epsilon_3}{2}\,.
\end{align}
\end{subequations}
Note that Eq.~\eqref{eq:solution-group-2} is reproduced for $\Omega=0$, as expected. The point groups of set IV can parametrize electromagnetic properties of uniaxial materials. For $\epsilon=\mathrm{diag}(\epsilon_t,\epsilon_t,\epsilon_l)$ and $\mu=\mathds{1}_3$, we find
\begin{subequations}
\label{eq:material-medium-example}
\begin{align}
\tilde{\kappa}^{00}&=\frac{1}{4}(\epsilon_l-3)+\frac{\epsilon_t}{2}\,,\quad k_{\mathrm{tr}}=4\tilde{\kappa}^{00}\,, \\[2ex]
\tilde{\kappa}^{11}&=\frac{1}{12}(5\epsilon_l-2\epsilon_t)-\frac{1}{4}\,,\quad
k^3=\frac{1}{6}(\epsilon_l-\epsilon_t)\,.
\end{align}
\end{subequations}
Each solution translates the electromagnetic parameters of a crystal into SME coefficients. Materials with a lesser symmetry group require at least 6 independent SME coefficients for parametrizing their permittivity and permeability. When symmetry increases, the number of independent coefficients decreases until the only one remaining for isotropic materials is $\tilde{\kappa}^{00}$. Note that $k_{\mathrm{tr}}=4\tilde{\kappa}^{00}$ applies to all materials, which is a consequence of the permeability being trivial and demonstrates that $k_{\mathrm{tr}}$ is indispensable, indeed. Since $\tilde{\kappa}^{00}$ is different for each material, $k_{\mathrm{tr}}$ differs, too.

Table~\ref{tab:example-materials} states explicit results for a handful of exemplary minerals. Datolite has a monoclinic unit cell with $\varphi_{xz}=90.15^{\circ}$. Since the latter value differs only slightly from $90^{\circ}$, the coefficients $\tilde{\kappa}^{13}$ and $k^6$ are heavily suppressed and omitted. Moreover, the materials shown in the first four lines are dominated by components of the first principal sector, and components of the second principal sector are strongly suppressed. A single independent component of the second principal sector only becomes important for the fifth example. For isotropic materials, the structure $\tilde{\kappa}=\tilde{\kappa}^{00}\mathrm{diag}(1,1/3,1/3,1/3)$ is evident, and the coefficients $k^a$ do not contribute at all.

Note that we have not been able to identify suitable magnetic compounds with permeability components $\gg 1$ that can be embedded in the current setting. Such materials usually have a nonlinear response to external magnetic fields in the form of a hysteresis, whereas we assumed that the electromagnetic responses are at least approximately linear.

We emphasize that the values of SME coefficients provided in Tab.~\ref{tab:example-materials} do not describe a violation of Lorentz invariance at the fundamental level. Here, each coefficient is a material parameter contributing to an effective description of electromagnetic properties of this material. After all, the SME is a proper framework for this endeavor. In principle, each SME coefficient could be supplemented with a label representing the material. We refrain from doing so, though, since this would make the notation unwieldy. As expected, the explicit values for many coefficients are of $\mathcal{O}(1)$, with some being $\gg 1$. Therefore, perturbative treatments, which have been quite successful for fundamental Lorentz violation, are not expected to yield meaningful results here. 

\subsection{Magnetoelectricity}

The magnetoelectric coupling of a material is parametrized by either the components $\alpha_{ij}$ of Eq.~\eqref{eq:constitutive-relations-SME} or $\beta_{ij}$ of Eq.~\eqref{eq:constitutive-relations-response}. The symmetry of the atomic lattice, which is encoded in its magnetic point group, dictates the structure of $\alpha$ and $\beta$. The behavior under time reversal has a direct impact on the components of $\alpha,\beta$ and eliminates those that are not in accordance with the symmetry. This allows one to predict which materials exhibit magnetoelectric effects and what their preferred coupling directions are. Equations~\eqref{eq:constitutive-relations-response} and \eqref{eq:susceptibilities} already tell us how $\alpha$ and $\beta$ are governed by the SME quantity $k_F$. Therefore, symmetry operations can be applied to the latter, which reduces the number of independent components of~$k_F$.

The magnetoelectric couplings for all 122 magnetic point groups and 14 Curie limiting groups are determined from the characteristic symmetry operations of each group. Note that it is unnecessary to analyze all groups, since fundamental symmetry arguments simplify the investigation. Among the 122 magnetic point groups, 32 combine time reversal with any other symmetry operation. Atomic lattices described by these groups cannot lead to magnetoelectric effects. This is because T inverts the magnetic moment, which prevents electric polarization from coupling to the magnetic field. Thus, there remain 90 magnetic point groups that are compatible with magnetoelectricity. This includes groups that combine T with other symmetry operations such as $\mathbf{m}$' or $\mathbf{2}/\mathbf{m}$', as well as groups that lack the time reversal transformation T, e.g., $\mathbf{m}$.

We indent to introduce our procedure on the basis of $\beta$ in Eq.~\eqref{eq:constitutive-relations-response}. Resorting to tables of generators for irreducible representations and applying invariance conditions to $\beta$ optimizes our analysis of the remaining 90 groups. The nonzero SME coefficients then follow through interdependence relations. Note that the Curie limiting groups imply particularly simple matrix forms, while atomic lattices devoid of symmetries, such as triclinic systems, may permit all 9 independent components of $\beta$.

In the following, we have to understand how $\beta$ behaves under discrete symmetry operations. Let us consider coordinate transformations in 3 dimensions described by a $(3\times 3)$ matrix $\mathcal{D}$. They can be proper or improper rotations, such as reflections and spatial inversions, respectively. The former preserve the handedness of the coordinate system, whereas the latter change it such that $\det(\mathcal{D})=\pm 1$. We start with the fundamental transformation law for an axial vector with components $B_i$:
\begin{equation}
B_{i}=\pm\det(\mathcal{D})\mathcal{D}_{ij}B_{j}\,,
\end{equation}
where the signs indicate conservation and violation of T symmetry, respectively. Magnetization transforms as an axial vector,
\begin{subequations}
\begin{align}
M_{i}'  &  =\pm\det(\mathcal{D})\mathcal{D}_{ij}M_{j}=\pm\det(\mathcal{D})\mathcal{D}_{ij}\beta_{kj}E_{k} \nonumber\\
&=\pm \det(\mathcal{D})\mathcal{D}_{ij}\beta_{kj}\mathcal{D}_{lk}E_{l}:=\beta'_{li}E_{l}'\,, \\[1ex]
\beta'_{li} & =\pm\det(\mathcal{D})\mathcal{D}_{ij}\beta_{kj}\mathcal{D}_{lk}\,,
\end{align}
\end{subequations}
whereupon $\beta$ inherits its transformation properties from those of the magnetization and magnetic field.

Since $\kappa_{HB}$ is symmetric and polar, $\beta$ transforms as an axial tensor, so does $\kappa_{DB}$; cf.~Eq.~\eqref{eq:susceptibilities}. The symmetry transformations of the magnetic point groups govern the tensor $\beta$, as it relates polarization to magnetization. The transformations for all noncubic groups can be applied directly to the matrix $\kappa_{DB}$. On the contrary, for cubic systems, the operations may eliminate each component of $\kappa_{DB}$. To describe cubic magnetoelectric crystals within the SME, the $\theta$ term of Eq.~\eqref{eq:lagrangian-theta} has to step in. According to Eq.~\eqref{eq:constitutive-properties-CFJ}, this setting will be more than suitable for such materials.
\begin{table*}[t]
\centering
\begin{tabular}{cccc} 
\toprule
Set & Magnetic Point Groups & \# Coeffs. & Allowed form of $\alpha$ of Eq.~\eqref{eq:definition-beta} \\
\midrule
I & \begin{minipage}{5cm}
\centering
\textbf{1}, $\bar{\mathbf{1}}$' \\
\end{minipage} & $3+5+1$ &
$\begin{pmatrix}
2(\theta+k^{2}) & -\tilde{\kappa}^{03} & \tilde{\kappa}^{02}\\
\tilde{\kappa}^{03}-2k^{9} & 2(\theta-k^{1}) & -\tilde{\kappa}^{01}\\
-\tilde{\kappa}^{02}+2k^{8} & \tilde{\kappa}^{01}+2k^{10} & 2(\theta+k^{1}-k^{2}) \\
\end{pmatrix}$ \\[0.5cm]
II & \begin{minipage}{5cm}
\centering
$\mathbf{2}$, $\mathbf{m}$', $\mathbf{2}/\mathbf{m}$'
\end{minipage} & $1+3+1$ &
$\begin{pmatrix}
2(\theta+k^{2}) & 0 & \tilde{\kappa}^{02}\\
0 & 2(\theta-k^{1}) & 0\\
-\tilde{\kappa}^{02}+2k^{8} & 0 & 2(\theta+k^{1}-k^{2}) \\
\end{pmatrix}$ \\[0.5cm]
III & \begin{minipage}{5cm}
\centering
$\mathbf{2}$', $\mathbf{m}$, $\mathbf{2}$'/$\mathbf{m}$ \\
\end{minipage} & $2+2+0$ &
$\begin{pmatrix}
0 & -\tilde{\kappa}^{03} & 0\\
\tilde{\kappa}^{03}-2k^{9} & 0 & -\tilde{\kappa}^{01}\\
0 & \tilde{\kappa}^{01}+2k^{10} & 0%
\end{pmatrix}$ \\[0.5cm]
\begin{minipage}{1cm}
\centering
\vspace{0.3cm}IV
\end{minipage} & \begin{minipage}{5cm}
\centering
\vspace{0.3cm}
$\mathbf{222}$, $\mathbf{m}$'$\mathbf{m}$'$\mathbf{2}$, $\mathbf{m}$'$\mathbf{m}$'$\mathbf{m}$' \\
\end{minipage} & \begin{minipage}{1.5cm}
\centering
\vspace{0.3cm}$0+2+1$
\end{minipage} & \begin{minipage}{5cm}
\centering
\vspace{0.3cm}
$2\theta\mathds{1}_3+2\,\mathrm{diag}(k^{2},-k^{1},k^{1}-k^{2})$
\end{minipage} \\[0.5cm]
V & \begin{minipage}{5cm}
\centering
$\mathbf{2}$$\mathbf{2}$'$\mathbf{2}$', $\mathbf{2mm}$, $\mathbf{m}$'$\mathbf{m2}$', $\mathbf{m}$'$\mathbf{mm}$
\end{minipage} & $1+1+0$ &
$\begin{pmatrix}
0 & 0 & 0\\
0 & 0 & -\tilde{\kappa}^{01}\\
0 & 2k^{10}+\tilde{\kappa}^{01} & 0%
\end{pmatrix}$ \\[0.5cm]
VI & \begin{minipage}{6cm}
\centering
\textbf{3}, $\bar{\mathbf{3}}$', \textbf{4}, $\bar{\mathbf{4}}$', $\mathbf{4}$/$\mathbf{m}$', \textbf{6}, $\bar{\mathbf{6}}$', \textbf{6}/$\mathbf{m}$', $\boldsymbol{\infty}$, $\boldsymbol{\infty}\mathbf{m}$'
\end{minipage} & $1+1+1$ &
$\begin{pmatrix}
2(\theta+k^2) & -\tilde{\kappa}^{03} & 0 \\
\tilde{\kappa}^{03} & 2(\theta+k^2) & 0 \\
0 & 0 & 2(\theta-2k^2) \\
\end{pmatrix}$ \\[0.5cm]
VII & \begin{minipage}{5cm}
\centering $\mathbf{4}$', $\bar{\mathbf{4}}$, $\mathbf{4}$'/$\mathbf{m}$'
\end{minipage} & $1+1+0$ &
$\begin{pmatrix}
2k^{1} & -\tilde{\kappa}^{03} & 0\\
-\tilde{\kappa}^{03} & -2k^{1} & 0\\
0 & 0 & 0%
\end{pmatrix}$ \\[0.5cm]
\begin{minipage}{1cm}
\centering
\vspace{0.3cm}VIII
\end{minipage} & \begin{minipage}{6cm}
\vspace{0.3cm}
$\mathbf{4}$'$\mathbf{22}$, $\mathbf{4}$'$\mathbf{mm}$', $\bar{\mathbf{4}}$$\mathbf{2m}$, $\mathbf{\bar{4}2}$'$\mathbf{m}$', $\mathbf{4}$'/$\mathbf{m}$'$\mathbf{mm}$'
\end{minipage} & \begin{minipage}{1.5cm}
\centering
\vspace{0.3cm}$0+1+0$
\end{minipage} &
\begin{minipage}{6cm}
\centering
\vspace{0.3cm}
$2k^{1}\,\mathrm{diag}(1,-1,0)$
\end{minipage} \\[0.5cm]
IX & \begin{minipage}{7cm}
\centering
$\mathbf{32}$', $\mathbf{3m}$, $\mathbf{\bar{3}}$'$\mathbf{m}$, $\mathbf{42}$'$\mathbf{2}$', $\mathbf{4mm}$, $\mathbf{\bar{4}}$'$\mathbf{2}$'$\mathbf{m}$, $\mathbf{4}$/$\mathbf{m}$'$\mathbf{mm}$, \\[0.5ex]
$\mathbf{62}$'$\mathbf{2}$', $\mathbf{6mm}$, $\mathbf{\bar{6}}$'$\mathbf{m2}$', $\mathbf{6}$/$\mathbf{m}$'$\mathbf{mm}$, $\boldsymbol{\infty}$$\mathbf{2}$', $\boldsymbol{\infty}$/$\mathbf{m}$'$\mathbf{m}$
\end{minipage} & $1+0+0$ &
$\tilde{\kappa}^{03}\begin{pmatrix}
0 & -1 & 0\\
1 & 0 & 0\\
0 & 0 & 0%
\end{pmatrix}$ \\[0.5cm]
\begin{minipage}{1cm}
\centering
\vspace{0.2cm}X
\end{minipage} & \begin{minipage}{8cm}
\vspace{0.2cm}
$\mathbf{32}$, $\mathbf{3m}$', $\mathbf{\bar{3}}$'$\mathbf{m}$', $\mathbf{422}$, $\mathbf{4m}$'$\mathbf{m}$', $\mathbf{\bar{4}}$'$\mathbf{2m}$', $\mathbf{4}$/$\mathbf{m}$'$\mathbf{m}$'$\mathbf{m}$', $\mathbf{622}$, \\[0.5ex]
$\mathbf{6m}$'$\mathbf{m}$', $\mathbf{\bar{6}}$'$\mathbf{m}$'$\mathbf{2}$, $\mathbf{6}$/$\mathbf{m}$'$\mathbf{m}$'$\mathbf{m}$', $\boldsymbol{\infty}$$\mathbf{2}$, $\boldsymbol{\infty}$/$\mathbf{m}$'$\mathbf{m}$' \\
\end{minipage} & \begin{minipage}{1.5cm}
\centering
\vspace{0.2cm}$0+1+1$
\end{minipage} &
\begin{minipage}{6cm}
\centering
\vspace{0.2cm}$2\theta\mathds{1}_3+2k^{2}\,\mathrm{diag}(1,1,-2)$
\end{minipage} \\[0.5cm]
\begin{minipage}{1cm}
\centering
\vspace{0.3cm}XI
\end{minipage} & \begin{minipage}{8cm}
\centering
\vspace{0.3cm}
\textbf{23}, $\mathbf{m}$'\textbf{3}, \textbf{432}, $\bar{\mathbf{4}}$'$\mathbf{3m}$', $\textbf{m}$'$\mathbf{3m}$', $\boldsymbol{\infty}\boldsymbol{\infty}$, $\boldsymbol{\infty\infty}\mathbf{m}$' \\
\end{minipage} & \begin{minipage}{1.5cm}
\centering
\vspace{0.3cm}
$0+0+1$
\end{minipage} & \begin{minipage}{3cm}
\centering
\vspace{0.3cm}
$2\theta\mathds{1}_3$
\end{minipage} \\[0.5cm]
XII & other & $0+0+0$ &
$\begin{pmatrix}
0 & 0 & 0\\
0 & 0 & 0\\
0 & 0 & 0%
\end{pmatrix}$ \\
\bottomrule
\end{tabular}
\caption{Forms of the SME magnetoelectric-coupling tensor $\kappa_{DB}$ consistent with magnetic point groups. The first column runs through the different sets and the second lists the magnetic point groups. The third column provides the numbers of nonzero independent coefficients of the first and second principal sectors of $k_F$, as well as the single coefficient $\theta$, if present. Last but not least, the allowed form of $\kappa_{DB}$ can be found in the fourth column.}
\label{tab:magnetoelectric}
\end{table*}

\subsubsection{Implementation of magnetic point groups}

Equations~\eqref{eq:kappa-DB-sector-1} and \eqref{eq:kappa-DB-sector-2} tell us that the generic matrix $\kappa_{DB}$ containing degrees of freedom from both principal sectors of $k_F$ is of the form
\begin{equation}
\label{eq:total-kappa-DB}
\kappa_{DB}=\left(\begin{array}[c]{ccc}%
2(\theta+k^{2}) & -\tilde{\kappa}^{03} & \tilde{\kappa}^{02}\\
\tilde{\kappa}^{03}-2k^{9} & 2(\theta-k^{1}) & -\tilde{\kappa}^{01}\\
-\tilde{\kappa}^{02}+2k^{8} & \tilde{\kappa}^{01}+2k^{10} & 2(\theta+k^{1}-k^{2}) \\
\end{array}
\right)\,.
\end{equation}
Recall that $\kappa_{DB}$ is an axial tensor. The transformation law of the SME magnetoelectric coupling $\kappa_{DB}$ under rotations is analogous to that of $\kappa_{DE}$. For a generic rotation with angle $\Omega$ around $[hkl]$ it holds that
\begin{equation}
\label{eq:rotation-kappa-DB}
(\kappa_{DB}')^{ij}=\mathcal{R}^{[hkl]}_{il}(\Omega)(\kappa_{DB})^{lm}\mathcal{R}^{[hkl]}_{jm}(\Omega)\,,
\end{equation}
as $\det(\mathcal{R}^{[hkl]})=1$. However, if the latter is subject to a reflection, it transforms as
\begin{equation}
\label{eq:reflection-kappa-DB}
(\kappa_{DB}'')^{ij}=-\mathcal{M}^{[hkl]}_{il}(\Omega)(\kappa_{DB})^{lm}\mathcal{M}^{[hkl]}_{jm}(\Omega)\,,
\end{equation}
due to $\det(\mathcal{M}^{[hkl]})=-1$. Under time reversal, the entire matrix $\kappa_{DB}$ changes sign.

We start with the crystallographic group $\mathbf{1}$ as well as the magnetic point group $\mathbf{\bar{1}}$'. Both have a simple structure that does not restrict the degrees of freedom of Eq.~\eqref{eq:total-kappa-DB} in any way; see the first line of Tab.~\ref{tab:magnetoelectric}. In fact, the effects of the parity and time reversal transformations in $\mathbf{\bar{1}}$' cancel each other.

Let us demonstrate the procedure for the two nontrivial cases of the magnetic point groups $\mathbf{2}$ and $\boldsymbol{\infty}\mathbf{m}$'. The former is a low-symmetry group governing a monoclinic crystal. A rotation by $\pi$ is performed around the crystallographic direction [010], which is the principal symmetry direction in monoclinic crystal systems. Based on Eq.~\eqref{eq:rotation-kappa-DB}, this symmetry transformation implies the following transformed SME magnetoelectric coupling:
\begin{subequations}
\begin{equation}
(\kappa_{DB}')^{ij}=\mathcal{R}^{[010]}_{il}(\pi)(\kappa_{DB})^{lm}\mathcal{R}^{[010]}_{jm}(\pi)\,,
\end{equation}
\begin{equation}
\kappa_{DB}'=\begin{pmatrix}
2(\theta+k^2) & \tilde{\kappa}^{03} & \tilde{\kappa}^{02} \\
-\tilde{\kappa}^{03}+2k^9 & 2(\theta-k^1) & \tilde{\kappa}^{01} \\
-\tilde{\kappa}^{02}+2k^8 & -(\tilde{\kappa}^{01}+2k^{10}) & 2(\theta+k^1-k^2) \\
\end{pmatrix}\,.
\end{equation}
\end{subequations}
Then, the invariance condition $\kappa_{DB}'=\kappa_{DB}$ gets rid of the coefficients $\tilde{\kappa}^{01}$, $\tilde{\kappa}^{03}$, $k^9$, and $k^{10}$. Therefore, the SME magnetoelectric-coupling tensor is reduced to
\begin{equation}
\kappa_{DB}=\left(
\begin{array}
[c]{ccc}%
2(\theta+k^{2}) & 0 & \tilde{\kappa}^{02} \\
0 & 2(\theta-k^{1}) & 0 \\
-\tilde{\kappa}^{02}+2k^{8} & 0 & 2(\theta+k^{1}-k^{2}) \\
\end{array}
\right)\,,
\end{equation}
see the second line of Tab.~\ref{tab:magnetoelectric}.

The magnetic point group $\boldsymbol{\infty}\textbf{m}$' describes a high-symmetry uniaxial crystal. The symmetry transformations consist of continuous rotations around an infinite-order principal axis combined with time-reversed reflections in a plane containing that axis. For example, compatibility with rotations around $[001]$ imposes the restriction $k^1=-k^2$ and eliminates all remaining coefficients except~$\tilde{\kappa}^{03}$. The latter only learns its fate when performing reflections at a mirror that contains the $[001]$ direction. For invariance of $\kappa_{DB}$, there is no choice other than $\tilde{\kappa}^{03}=0$. We then end up with a diagonal-matrix representation for $\kappa_{DB}$, where the first two entries are equal:
\begin{equation}
\kappa_{DB}=2\theta\mathds{1}_3+2k^{2}\,\mathrm{diag}(1,1,-2)\,,
\end{equation}
see the tenth line in Tab.~\ref{tab:magnetoelectric}. Considering the axes $[100]$ and $[010]$, respectively, leads to corresponding results.

In an analogous way, the procedure described is extended to all 90 magnetic point groups and 14 Curie limiting groups. Note that besides groups with pure time-reversal symmetry, there are further groups that cannot support magnetoelectric effects. Out of the original 90 groups, only 58 viable groups remain. Doing so complements Tab.~\ref{tab:magnetoelectric}, which lists all possible matrices $\kappa_{DB}$ whose forms are compatible with symmetry transformations of magnetic point groups.

Several comments are in order. First, the complexity of $\kappa_{DB}$ reduces with increasing symmetry. The more symmetry a magnetic material possesses, the more restricted the degrees of freedom are in the SME magnetoelectric coupling. Second, observing how the component coefficients are successively reduced, we can follow 3 different routes: I $\mapsto$ II $\mapsto$ IV $\mapsto$ VII/IX $\mapsto$ X or I $\mapsto$ III $\mapsto$ IV $\mapsto$ X or I $\mapsto$ VI $\mapsto$ VII/VIII $\mapsto$ X. Third, the number of component coefficients of $\kappa_{DB}$ is always even, unless only a single one remains. Fourth, all the coefficients except $k^8,k^9,k^{10}$ are multiplied by $\mathfrak{so}(3)$ generators. The latter coefficients are eliminated first when more symmetries are imposed.

Table~\ref{tab:magnetoelectric} emphasizes again that modified Maxwell theory does not account for highly symmetric cubic crystals. The tracelessness of $\kappa_{DB}$ in combination with the high symmetry of cubic systems, e.g., groups like \textbf{23}, \textbf{432}, $\textbf{m}$'$\mathbf{3}$, and $\boldsymbol{\infty}\boldsymbol{\infty}$ eliminates all components of $\kappa_{DB}$. Their symmetry requires a nonzero isotropic component. As mentioned before, the $\theta$ term of Sec.~\ref{eq:CFJ-theory} comes to a rescue. In fact, it is indispensable for describing magnetoelectricity for such materials.

In an analogous way, the magnetoelectric coupling $\beta$ of Eq.~\eqref{eq:susceptibilities} can be investigated. In contrast to $\kappa_{DB}$, the quantity $\beta$ is not traceless. This difference is due to the additional degrees of freedom of $\kappa_{HB}$, which is not necessarily traceless when taking $k_{\mathrm{tr}}$ into account. While $\kappa_{DB}$ violates the dual symmetry of electromagnetism, $\beta$ depends on both $\kappa_{DB}$ and $\kappa_{HB}$, whereupon its trace can be nonzero in suitable systems.

\subsubsection{Application to real magnetoelectric materials}

Let us consider an uniaxial material with nontrivial permittivity and magnetoelectric coupling described by the following diagonal matrices:
\begin{equation}
\label{eq:properties-magnetoelectric-material}
\epsilon=\mathrm{diag}(\epsilon_t,\epsilon_t,\epsilon_l)\,,\quad \alpha=\mathrm{diag}(\alpha_t,\alpha_t,\alpha_l)\,,
\end{equation}
where the permeability is trivial: $\mu=\mathds{1}_3$. So there are 6 nontrivial quantities: a transverse permittivity $\epsilon_t$ and a transverse magnetoelectric coupling $\alpha_t$, which occur two-fold, a longitudinal permittivity $\epsilon_l$, and a longitudinal magnetoelectric coupling $\alpha_l$.

As an explicit example, we consult a crystal of magnetic point group $\bar{\mathbf{3}}$'$\mathbf{m}$' such as the compound $\mathrm{Cr_2O_3}$, which is a well-known magnetoelectric material~\cite{Sihvola:1995,Hehl:2007jy,Hehl:2007ut} that bears the name Eskolaite. According to Tabs.~\ref{tab:electric_susceptibility}, \ref{tab:magnetic_susceptibility}, and \ref{tab:magnetoelectric}, we propose characteristic SME tensors of the form
\begin{subequations}
\begin{align}
\kappa_{DE}&=\mathrm{diag}(\tilde{\kappa}^{00}-\tilde{\kappa}^{11},\tilde{\kappa}^{00}-\tilde{\kappa}^{11},2\tilde{\kappa}^{11})+\frac{k_{\mathrm{tr}}}{6}\mathds{1}_3\,, \displaybreak[0]\\[1ex]
\kappa_{HB}&=-\kappa_{DE}+\frac{k_{\mathrm{tr}}}{3}\mathds{1}_3-2\,\mathrm{diag}\left(k^3,k^3,-2k^3\right)\,, \displaybreak[0]\\[1ex]
\kappa_{DB}&=2\theta\mathds{1}_3+2k^2\,\mathrm{diag}(1,1,-2)\,,
\end{align}
\end{subequations}
where $\theta$ parametrizes a pseudoscalar part according to Eq.~\eqref{eq:lagrangian-theta}. The electromagnetic tensors are given by Eqs.~\eqref{eq:permittivity}, \eqref{eq:magnetoelectric-coupling}. Setting the latter equal to those of the material, Eq.~\eqref{eq:properties-magnetoelectric-material}, leads to a linear system of 6 equations for the 6 SME coefficients, i.e., the number of coefficients matches the number of nontrivial electromagnetic quantities, as expected. Solving this system provides
\begin{subequations}
\begin{align}
\tilde{\kappa}^{00}&=\frac{1}{4}(\mathrm{tr}(\epsilon)-3)\,,\quad \tilde{\kappa}^{11}=\frac{1}{12}(5\epsilon_l-2\epsilon_t-3)\,, \displaybreak[0]\\[1ex]
k^2&=\frac{1}{6}(\alpha_t-\alpha_l)\,,\quad k^3=\frac{1}{6}(\epsilon_l-\epsilon_t)\,, \displaybreak[0]\\[1ex]
k_{\mathrm{tr}}&=4\tilde{\kappa}^{00}\,,\quad \theta=\frac{\mathrm{tr}(\alpha)}{6}\,.
\end{align}
\end{subequations}
The solution tells us that $\tilde{\kappa}^{00}$, $\tilde{\kappa}^{11}$, $k_{\mathrm{tr}}$, and $k^3$ are linked to the nontrivial permittivity, whereas $k^2$ and $\theta$ are directly related to the nontrivial magnetoelectric properties; cf.~Eq.~\eqref{eq:material-medium-example}. Explicit numbers of the coefficients for this material are compiled in the last line of Tab.~\ref{tab:example-materials}.

Note that the permittivity and magnetoelectric-coupling tensors of $\mathrm{Cr_2O_3}$ are simultaneously diagonal. Physically, this means that the principal axes of both electromagnetic properties correspond to each other. An intriguing question is whether the crystallographic structure necessarily implies that this is the case. For example, according to Tabs.~\ref{tab:electric_susceptibility}, \ref{tab:magnetic_susceptibility}, and \ref{tab:magnetoelectric}, the group \textbf{222} is associated with diagonal $\kappa_{DE}$, $\kappa_{HB}$, and $\kappa_{DB}$, respectively. However, when taking the T transformation into account via \textbf{2}\textbf{2}'\textbf{2}', the corresponding $\kappa_{DB}$ is nondiagonal.

\section{Birefringence in the SME}
\label{eq:birefringence}

Birefringence is an intriguing effect that some specific crystals exhibit. A birefringent crystal duplicates an object that is situated behind the crystal. These two images are caused by two propagation modes associated with two different refractive indices. One refractive index is isotropic, whereas the second is direction-dependent. The mode related to the isotropic refractive index behaves like a ray in any nonbirefringent crystal. It obeys Snell's law and is called the ordinary ray. Contrarily, the second mode is anisotropic and does not propagate according to Snell's law. In particular, upon orthogonal incidence on the crystal surface, the extraordinary ray is refracted. Therefore, any object seen through the crystal appears twice.

At the microscopic level, birefringence is a consequence of preferred molecular alignments, which implies a direction-dependent polarizability. These directions are known as optical axes. There are crystals with a single optical axis, which are called uniaxial. Biaxial crystals exhibit two optical axes. Interestingly, birefringence is directly related to other effects such as optical activity~\cite{Barron:2009}.

As a generic parametrization for the electromagnetic properties of material media, the electromagnetic sector of the SME is capable of describing birefringence. However, we will notice that there are crucial differences between birefringence in crystals and birefringence as parametrized in the SME. We have already seen that the CPT-even part of the minimal electromagnetic SME decomposes into two principal sectors; see Sec.~\ref{sec:principal-sectors}. Birefringence occurs in both. In the first sector, it is of second order in the SME coefficients, whereas it is of first order in the second sector. In the context of vacuum birefringence, the first sector does not play a role. Any phenomenological study of a hypothetical fundamental birefringence in nature refers to the second sector.

Birefringence \textit{in vacuo} has been tightly constrained. An analysis of the CMB polarization constrains coefficients of CFJ theory down to $\unit[10^{-44}]{GeV}$ -- $\unit[10^{-43}]{GeV}$~\cite{Carroll:1989vb,Kostelecky:2008ts}, which is many orders of magnitude smaller than the lower limit of the photon mass. Spectropolarimetry leads to bounds on dimensionless coefficients $k^a$ that lie in the ballpark of $10^{-38}$ -- $10^{-34}$~\cite{Kostelecky:2008ts}. The main conclusion is that fundamental birefringence, if it exists in nature, is excessively suppressed.

Our interest here is to describe birefringence in optical media via the electromagnetic sector of the SME. Since we will work at all orders in the SME coefficients, the forthcoming study will reveal certain characteristics of birefringence in the SME that have not been pointed out in the contemporary literature. The generic dispersion equation for electromagnetic waves in the SME is involved but can be expressed in closed form as follows~\cite{Kostelecky:2009zp}:
\begin{subequations}
\label{eq:quartic-photon-dispersion-equation}
\begin{align}
0&=\mathcal{G}^{\mu\nu\varrho\sigma}p_{\mu}p_{\nu}p_{\varrho}p_{\sigma}\,, \\[1ex]
\mathcal{G}^{\mu\nu\varrho\sigma}&=\frac{1}{4!}\varepsilon_{\alpha\beta\gamma\delta}\varepsilon_{\zeta\eta\kappa\lambda}\chi^{\alpha\beta\zeta(\mu}\chi^{\nu|\gamma\eta|\varrho}\chi^{\sigma)\delta\kappa\lambda}\,, \\[1ex]
\label{eq:electromagnetic-response}
\chi^{\mu\nu\varrho\sigma}&=\frac{1}{2}(\eta^{\mu\varrho}\eta^{\nu\sigma}-\eta^{\nu\varrho}\eta^{\mu\sigma})+(k_F)^{\mu\nu\varrho\sigma}\,.
\end{align}
\end{subequations}
Here, $p_{\mu}$ is the covariant wave four-vector, $\chi^{\mu\nu\varrho\sigma}$ the electromagnetic response tensor, and $\mathcal{G}^{\mu\nu\varrho\sigma}$ the Tamm-Rubilar tensor, which is expressed in terms of the Levi-Civita symbol $\varepsilon^{\mu\nu\varrho\sigma}$ in Minkowski spacetime. The expression is symmetrized over the indices enclosed by parentheses, but indices contained within pairs of vertical lines are excluded from symmetrization. Obukhov, Fukui, and Rubilar independently obtained an analogous result without resorting to the SME~\cite{Obukhov:2000nw}, modulo irrelevant prefactors. Actually, they are the ones who introduced $\mathcal{G}^{\mu\nu\varrho\sigma}$ in the form stated previously. It is enlightening to look at special cases of Eq.~\eqref{eq:quartic-photon-dispersion-equation}, which shall be done below.

\subsection{First principal sector}

The generic dispersion equation for the first principal sector follows from Eq.~\eqref{eq:quartic-photon-dispersion-equation} when resorting to the parametrization of Eq.~\eqref{eq:nonbirefringent-ansatz}. The tool \textit{Act} \cite{xTensor:2025} has proven highly valuable for evaluating the complicated tensor algebra. The dispersion equation amounts to
\begin{align}
\label{eq:dispersion-equation-sector-1}
0&=\left(1-\frac{1}{2}\tilde{\kappa}_{\mu\nu}\tilde{\kappa}^{\mu\nu}+\frac{1}{3}\tilde{\kappa}^{\mu}_{\phantom{\mu}\nu}\tilde{\kappa}^{\nu}_{\phantom{\nu}\varrho}\tilde{\kappa}^{\varrho}_{\phantom{\varrho}\mu}\right)p^4 \notag \\
&\phantom{{}={}}+(2p_{\mu}\tilde{\kappa}^{\mu\nu}p_{\nu}+p_{\mu}\tilde{\kappa}^{\mu}_{\phantom{\mu}\nu}\tilde{\kappa}^{\nu}_{\phantom{\nu}\varrho}p^{\varrho}-p_{\mu}\tilde{\kappa}^{\mu}_{\phantom{\nu}\nu}\tilde{\kappa}^{\nu}_{\phantom{\nu}\varrho}\tilde{\kappa}^{\varrho}_{\phantom{\varrho}\sigma}p^{\sigma})p^2 \notag \\
&\phantom{{}={}}+p_{\mu}\tilde{\kappa}^{\mu\nu}p_{\nu}(p_{\varrho}\tilde{\kappa}^{\varrho\sigma}p_{\sigma}+p_{\varrho}\tilde{\kappa}^{\varrho}_{\phantom{\varrho}\sigma}\tilde{\kappa}^{\sigma}_{\phantom{\sigma}\kappa}p^{\kappa})\,,
\end{align}
with the $(4\times 4)$ matrix $\tilde{\kappa}^{\mu\nu}$ of Eq.~\eqref{eq:nonbirefringent-ansatz}. It can also be rewritten as a polynomial equation in $p_0$. However, its form is very involved, which is why we opted for putting it into App.~\ref{app:formulas}. For a generic $\tilde{\kappa}_{\mu\nu}$, the polynomial is of fourth order in $p_0$ and does not necessarily factorize into quadratic polynomials.

Let us go through some special cases. For a purely spacelike $\tilde{\kappa}^{ij}$, Eq.~\eqref{eq:dispersion-equation-sector-1} collapses significantly and is expressed as
\begin{subequations}
\begin{align}
0&=\mathcal{D}_1\mathcal{D}_2+\Delta\,, \displaybreak[0]\\[2ex]
\mathcal{D}_1&=p^2+p^i\tilde{\kappa}^{ij}p^j\,, \displaybreak[0]\\[2ex]
\mathcal{D}_2&=\left(1-\frac{1}{2}\tilde{\kappa}^{ij}\tilde{\kappa}^{ij}\right)p^2+p^i\tilde{\kappa}^{ij}p^j-p^i\tilde{\kappa}^{ij}\tilde{\kappa}^{jk}p^k\,, \displaybreak[0]\\[1ex]
\Delta&=\frac{1}{3}p_0^2\tilde{\kappa}^{ij}\tilde{\kappa}^{jk}\tilde{\kappa}^{ki}p^2\,.
\end{align}
\end{subequations}
Notably, it decomposes into two quadratic dispersion equations when the quantity $\Delta$ vanishes. Configurations with $\Delta=0$ are characterized by two square-root dispersion relations, which differ from each other at second order in $\tilde{\kappa}^{ij}$:
\begin{subequations}
\label{eq:dispersion-equations-sector-1}
\begin{align}
p_0^{(+)}&=\sqrt{\mathbf{p}^2-p^i\tilde{\kappa}^{ij}p^j}\,, \\[2ex]
p_0^{(-)}&=\sqrt{\mathbf{p}^2-\frac{p^i\tilde{\kappa}^{ij}p^j-p^i\tilde{\kappa}^{ij}\tilde{\kappa}^{jk}p^k}{1-\tilde{\kappa}^{ij}\tilde{\kappa}^{ij}/2}}\,.
\end{align}
\end{subequations}
While this difference is suppressed for fundamental Lorentz violation, it is significant in describing birefringence in materials.
\begin{table}
\centering
\begin{tabular}{cp{2.2cm}p{2.6cm}p{2.4cm}p{0.01cm}}
\toprule
Case & \centering $v^{\mu}$ and $w^{\mu}$ & \centering Properties & \centering References & \\
\midrule
1 & \centering $(\sqrt{2\tilde{\kappa}_{\mathrm{tr}}},0,0,0)$ $(\sqrt{2\tilde{\kappa}_{\mathrm{tr}}},0,0,0)$ & \centering Isotropic, nonbirefringent & 
\centering\cite{Kaufhold:2007qd,Klinkhamer:2008ky,Hohensee:2008xz,Klinkhamer:2010zs,Klinkhamer:2011ez,Diaz:2016dpk,Duenkel:2021gkq,Duenkel:2021szq,Duenkel:2023nlk,Amram:2023jlc} & \\
2 & \centering $(0,0,0,\sqrt{2\tilde{k}^3})$ $(0,0,0,\sqrt{2\tilde{k}^3})$ & \centering Anisotropic, nonbirefringent & \centering\cite{Kaufhold:2007qd} & \\
3 & \centering $(1,0,0,0)$ $(0,\tilde{\kappa}^{01},\tilde{\kappa}^{02},\tilde{\kappa}^{03})$ & \centering Parity-odd, birefringent & \centering\cite{Kaufhold:2007qd,Schreck:2011ai,Bocquet:2010ke} & \\
\bottomrule
\end{tabular}
\caption{Special cases of Eq.~\eqref{eq:parameterization-2-vectors} considered in the literature in the context of fundamental Lorentz violation. Here, $\tilde{k}^3:=(2/3)\tilde{\kappa}^{33}$ in analogy to the definition of $\tilde{\kappa}_{\mathrm{tr}}$ of Eq.~\eqref{eq:definition-kappatr}.}
\label{tab:special-cases-sector-1}
\end{table}

Further interesting configurations of $\tilde{\kappa}^{\mu\nu}$ are parametrized in terms of two four-vectors $v^{\mu}$ and $w^{\mu}$:
\begin{equation}
\label{eq:parameterization-2-vectors}
\tilde{\kappa}^{\mu\nu}=\frac{1}{2}(v^{\mu}w^{\nu}+v^{\nu}w^{\mu})-\frac{1}{4}(v\cdot w)\eta^{\mu\nu}\,.
\end{equation}
For the latter, the dispersion equation~\eqref{eq:dispersion-equation-sector-1} manifestly factorizes into two independent quadratic dispersion equations \cite{Casana:2010nd}:
\begin{subequations}
\label{eq:dispersion-equations-sector-2}
\begin{align}
0&=\left(1-\frac{v\cdot w}{2}\right)p^2+(v\cdot p)(w\cdot p)\,, \\[1ex]
0&=\left(1-\frac{v^2w^2}{4}\right)p^2+(v\cdot p)(w\cdot p) \notag \\
&\phantom{{}={}}+\frac{1}{4}\left[v^2(w\cdot p)^2+w^2(v\cdot p)^2\right]\,.
\end{align}
\end{subequations}
Note that Eq.~\eqref{eq:parameterization-2-vectors} is not restricted to purely spacelike choices of $\tilde{\kappa}^{\mu\nu}$.
Birefringence does not necessarily occur for such a parametrization, which is not straightforward to see, though. For certain configurations of $v^{\mu}$ and $w^{\mu}$, the second dispersion equation contains the first, i.e., it is superfluous. Then, there is no birefringence.

Table~\ref{tab:special-cases-sector-1} lists some prominent examples for the parametrization of Eq.~\eqref{eq:dispersion-equations-sector-2} that have already been considered in fundamental Lorentz violation. 
The following example is beyond those of Tab.~\ref{tab:special-cases-sector-1} and parametrizes a purely spacelike $\tilde{\kappa}^{\mu\nu}$ with nonzero coefficient $\tilde{\kappa}^{12}$:
\begin{equation}
\label{eq:directions-k12}
(v^{\mu})=\begin{pmatrix}
0 \\
\sqrt{2\tilde{\kappa}^{12}} \\
0 \\
0 \\
\end{pmatrix}\,,\quad (w^{\mu})=\begin{pmatrix}
0 \\
0 \\
\sqrt{2\tilde{\kappa}^{12}} \\
0 \\
\end{pmatrix}\,,
\end{equation}
such that
\begin{subequations}
\label{eq:dispersions-birefringence-sector-2-k12}
\begin{align}
p_0^{(+)}&=\sqrt{p_1^2+p_2^2+p_3^2-2\tilde{\kappa}^{12}p_1p_2}\,, \\[1ex]
p_0^{(-)}&=\sqrt{\frac{p_1^2+p_2^2-2\tilde{\kappa}^{12}p_1p_2}{1-(\tilde{\kappa}^{12})^2}+p_3^2}\,.
\end{align}
\end{subequations}
Here, it is straightforward to perceive that the difference between both dispersion relations is of second order in the coefficient $\tilde{\kappa}^{12}$.
\begin{figure}
\centering
\includegraphics[scale=0.25]{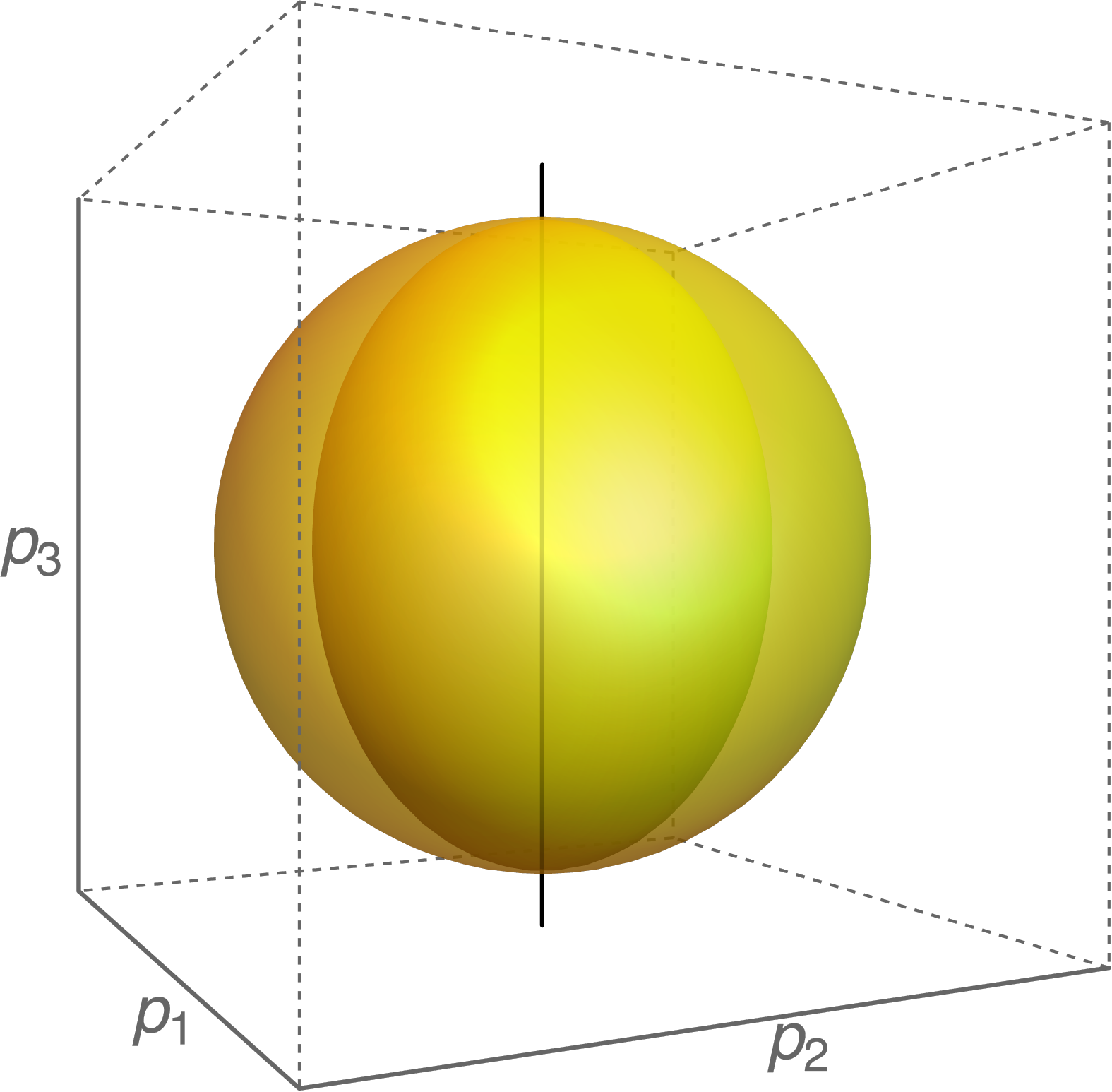}
\caption{Surface of constant $p^0$ based on Eq.~\eqref{eq:dispersion-equations-sector-2} for the parity-odd case of Tab.~\ref{tab:special-cases-sector-1}, where $\tilde{\kappa}^{01}=\tilde{\kappa}^{02}=0$ and $\tilde{\kappa}^{03}=1$. The black line indicates the optical axis.}
\label{fig:surface-kappa0i}
\end{figure}

According to the generic result of Eq.~\eqref{eq:dispersion-polynomial-sector-1}, there are configurations with intricate dispersion equations that contain linear or cubic terms in $p_0$. The resulting dispersion relations are often nontransparent, especially when they depend on cubic roots. We restrict our studies to dispersion relations expressed in terms of square roots only.

The optical axes correspond to propagation directions of electromagnetic waves along which both dispersion relations are the same. This implies the following condition for purely spacelike configurations of Eq.~\eqref{eq:dispersion-equations-sector-1}:
\begin{equation}
p^i\Big(2\tilde{\kappa}^{ik}\tilde{\kappa}^{kj}-\tilde{\kappa}^{kl}\tilde{\kappa}^{kl}\tilde{\kappa}^{ij}\Big)p^j=0\,.
\end{equation}
For example, for the dispersion relations of Eq.~\eqref{eq:dispersions-birefringence-sector-2-k12}, the latter requirement is satisfied for $p^1=p^2=0$, i.e., the optical axis points along the $z$ axis. Note that the optical axis is orthogonal to the spacelike parts of the four-vectors $v^{\mu}$ and $w^{\mu}$ of Eq.~\eqref{eq:directions-k12}. This holds for each purely spacelike coefficient, as can be verified explicitly. In general, each component $\tilde{\kappa}^{\mu\nu}$ is associated with a single optical axis, i.e., the SME coefficients of the first principal sector of $k_F$ describe uniaxial birefringence; see Tab.~\ref{tab:birefringence-sector-1}.
\begin{table}
\centering
\begin{tabular}{ccc}
\toprule
Component & Type & Optical axis \\
\midrule
$\tilde{\kappa}^{00}$ & Uniaxial & $\mathbf{e}_z$ \\
$\tilde{\kappa}^{11}$ & Uniaxial & $\mathbf{e}_y$ \\
$\tilde{\kappa}^{22}$ & Uniaxial & $\mathbf{e}_x$ \\
$\tilde{\kappa}^{0i}$ & Uniaxial & $\tilde{\kappa}^{0i}\mathbf{e}_i$ \\
$\tilde{\kappa}^{ij}$ ($i\neq j$) & Uniaxial & $\mathbf{e}_i\times\mathbf{e}_j$ \\
\midrule
\end{tabular}
\caption{Properties of birefringence for the first principal sector of $k_F$. The first column lists (sets of) component coefficients of $\tilde{\kappa}^{\mu\nu}$. The second refers to the birefringence type described and the third states the optical axis for each set. The symbol $\mathbf{e}_i$ denotes the $i$-th unit vector of the Cartesian basis in three dimensions.}
\label{tab:birefringence-sector-1}
\end{table}

For the purpose of illustration, consider Fig.~\ref{fig:surface-kappa0i} that shows the dispersion surface based on Eq.~\eqref{eq:dispersion-equations-sector-2} for the parity-odd case; see Tab.~\ref{tab:special-cases-sector-1}. The surface is a union of two ellipsoids touching each other at antipodal points. The latter are connected by a single optical axis, which is characteristic of the first principal sector.

\subsection{Second principal sector}

A possible parametrization of the second principal sector of $k_F$, which resembles Eq.~\eqref{eq:nonbirefringent-ansatz} for the first principal sector, reads
\begin{equation}
\label{eq:parameterization-second-principal-sector}
k_F^{\mu\nu\varrho\sigma}=\tilde{\kappa}_1^{\mu\varrho}\tilde{\kappa}_2^{\nu\sigma}-\tilde{\kappa}_1^{\mu\sigma}\tilde{\kappa}_2^{\nu\varrho}-\tilde{\kappa}_1^{\nu\varrho}\tilde{\kappa}_2^{\mu\sigma}+\tilde{\kappa}_1^{\nu\sigma}\tilde{\kappa}_2^{\mu\varrho}\,,
\end{equation}
with suitable symmetric $(4\times 4)$ matrices $\tilde{\kappa}_{1,2}$. The latter are tensor products
\begin{subequations}
\label{eq:kappas-second-principal-sector}
\begin{align}
\tilde{\kappa}_1&=\sum_{A,B} s_{AB}\xi^{(A)}\xi^{(B)}\,, \\[1ex]
\tilde{\kappa}_2&=k^a\sum_{A,B} t_{AB}\xi^{(A)}\xi^{(B)}\,,
\end{align}
\end{subequations}
where $s_{AB},t_{AB}\in \{0,\pm 1\}$, $k^a$ is an SME coefficient of the second principal sector, and $\xi^{(A,B)}$ with labels $A,B\in\{0\dots 3\}$ correspond to four-vectors of an orthonormal basis of Minkowski spacetime: $\xi^{(A)}_{\mu}=\delta^{A}_{\phantom{A}\mu}$.
Table~\ref{tab:parametrization-second-principal-sector} provides the actual expressions for each choice of $k^a$.
\begin{table}
\begin{tabular}{ccr}
\toprule
Coefficient $k^a$ & $\tilde{\kappa}_1$ & \multicolumn{1}{c}{$\tilde{\kappa}_2$} \\
\midrule
$k^1$ & $\xi^{(0)}\xi^{(1)}+\xi^{(1)}\xi^{(0)}$ & $\xi^{(2)}\xi^{(3)}+\xi^{(3)}\xi^{(2)}$ \\
$k^2$ & $\xi^{(0)}\xi^{(2)}+\xi^{(2)}\xi^{(0)}$ & $\xi^{(1)}\xi^{(3)}+\xi^{(3)}\xi^{(1)}$ \\
\midrule
$k^7$ & $\xi^{(1)}\xi^{(1)}$ & $\xi^{(2)}\xi^{(3)}+\xi^{(3)}\xi^{(2)}$ \\
\midrule
$k^4$ & $\xi^{(2)}\xi^{(2)}$ & $-\xi^{(1)}\xi^{(1)}+\xi^{(3)}\xi^{(3)}$ \\
$k^6$ &                                        & $\xi^{(1)}\xi^{(3)}+\xi^{(3)}\xi^{(1)}$ \\
$k^9$ &                                        & $\xi^{(0)}\xi^{(3)}+\xi^{(3)}\xi^{(0)}$ \\
\midrule
$k^3$ & $\xi^{(3)}\xi^{(3)}$ & $-\xi^{(1)}\xi^{(1)}+\xi^{(2)}\xi^{(2)}$ \\
$k^5$                   &                      & $\xi^{(1)}\xi^{(2)}+\xi^{(2)}\xi^{(1)}$ \\
$k^8$                   &                      & $\xi^{(0)}\xi^{(2)}+\xi^{(2)}\xi^{(0)}$ \\
$k^{10}$                &                      & $-\xi^{(0)}\xi^{(1)}-\xi^{(1)}\xi^{(0)}$ \\
\bottomrule
\end{tabular}
\caption{Parametrization of the second principle sector of $k_F$ according to Eq.~\eqref{eq:parameterization-second-principal-sector} with $\tilde{\kappa}_1$, $\tilde{\kappa}_2$, and $k^a$ given in Eq.~\eqref{eq:kappas-second-principal-sector}. The Lorentz indices are omitted, for brevity.}
\label{tab:parametrization-second-principal-sector}
\end{table}

\subsubsection{Dispersion equations}

The dispersion equation for the second principal sector is manifestly quartic and does not factorize. In all its generality, it is even more complicated than Eq.~\eqref{eq:dispersion-equation-sector-1}, which is why we omit it. The dispersion equations for each coefficient separately are short enough to be stated and already contain much information.

Let us start with the simplest cases that are those of $k^{3}\dots k^7$. The dispersion equation is of the generic form
\begin{equation}
\label{eq:dispersion-equation-k34567}
\tilde{\zeta}_{ij}p'^ip'^j\mathbf{p}'^2-2\tilde{\psi}_{ij}p'^ip'^j(p^0)^2+(p^0)^4=0\,,
\end{equation}
in terms of the wave vector $\mathbf{p}'$ and with symmetric $(3\times 3)$ matrices $\tilde{\zeta}$ and $\tilde{\psi}$. These are diagonal for $k^3$ and $k^4$, but have off-diagonal entries for $k^5\dots k^7$. For the latter coefficients, a coordinate rotation to the system of principal axes can be carried out. Then, the dispersion equation takes a more straightforward form:
\begin{equation}
\label{eq:dispersion-equation-k34567-transformed}
\zeta_i(p^i)^2\mathbf{p}^2-2\psi_i(p^i)^2(p^0)^2+(p^0)^4=0\,,
\end{equation}
expressed in terms of the transformed wave vector $\mathbf{p}$, where $\mathbf{p}'^2=\mathbf{p}^2$ is invariant. Here, the parameters $\zeta_i$ and $\psi_i$ are now vector-valued, and their components for each of the SME coefficients are stated in Tab.~\ref{tab:matrix-components-k34567}.
\begin{figure}
\centering
\includegraphics[scale=0.25]{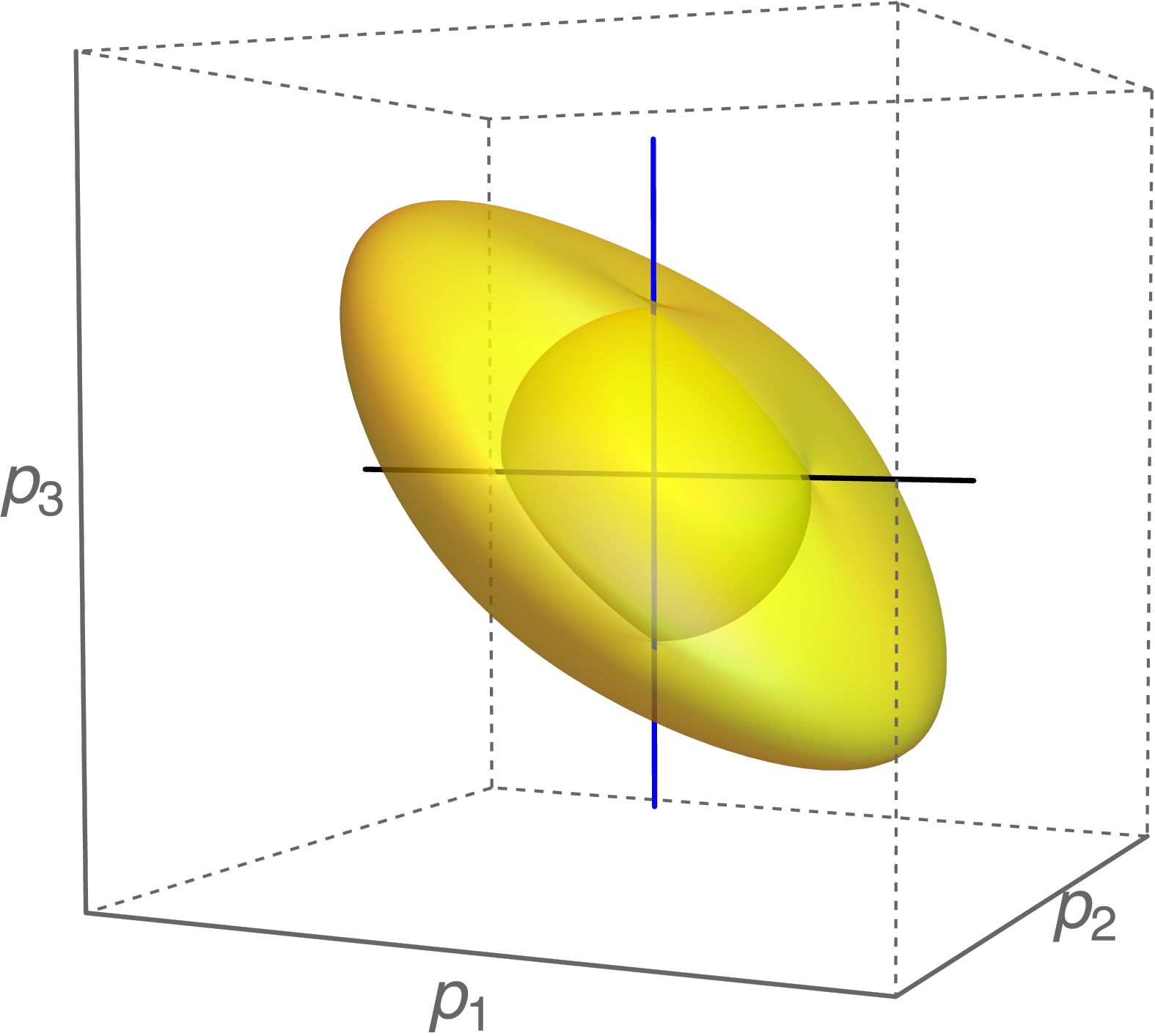}
\caption{Surface of constant $p^0$ from Eq.~\eqref{eq:dispersion-equation-k34567} for $k^6$. The two optical axes are color-coded in black and blue, respectively. Note the presence of two shells that intersect each other at four singular points. The black and blue lines depict the optical axes.}
\label{fig:surface-k6}
\end{figure}

Equation~\eqref{eq:dispersion-equation-k34567-transformed} for constant $p^0$ is known as Fresnel's wave surface. It holds for biaxial media and has a characteristic form, shown in Fig.~\ref{fig:surface-k6} for $k^6$. The surface consists of an outer and an inner shell. Both shells intersect at four points, which are singular, i.e., there is no tangent plane at each of the points. In fact, the wave vectors associated with the singular points all lie in a single plane. There are two pairs of antiparallel wave vectors whose directions correspond to the optical axes of the medium, which will be determined later.

According to Tab.~\ref{tab:matrix-components-k34567}, the dispersion equations in the principal coordinate system have the same functional form for $k^5\dots k^7$. In general, the dispersion equations for $k^3$, $k^4$, and $k^{5,6,7}$ are similar, where the parameters $\zeta_i$ and $\psi_i$ are simply permuted.

\begin{table}
\begin{tabular}{cc}
\toprule
$k$ & Nonzero parameters of Eqs.~\eqref{eq:dispersion-equation-k34567-transformed} \\
\midrule
$k^3$ & $\zeta_1=1-2k^3$, $\zeta_2=1+2k^3$, $\zeta_3=1-4(k^3)^2$ \\
      & $\psi_1=1-k^3$, $\psi_2=1+k^3$, $\psi_3=1$ \\
\midrule
$k^4$ & $\zeta_1=1-2k^4$, $\zeta_2=1-4(k^4)^2$, $\zeta_3=1+2k^4$ \\
      & $\psi_1=1-k^4$, $\psi_2=1$, $\psi_3=1+k^4$ \\
\midrule
$k^{5,6,7}$ & $\zeta_1=1-4(k^a)^2$, $\zeta_2=1-2k^a$, $\zeta_3=1+2k^a$ \\
            & $\psi_1=1$, $\psi_2=1-k^a$, $\psi_3=1+k^a$ \\
\bottomrule
\end{tabular}
\caption{Vector components to be inserted into Eq.~\eqref{eq:dispersion-equation-k34567-transformed} for each of the SME coefficients $k^3\dots k^7$.}
\label{tab:matrix-components-k34567}
\end{table}

The dispersion equations for the remaining coefficients are more complicated. Let us first of all focus on $k^1$ and $k^2$. For these cases, the dispersion equations can, indeed, be cast into the form
\begin{subequations}
\label{eq:dispersion-equation-k12}
\begin{equation}
K(p^1,p^2,p^3,\mathrm{i}p^0)=0\,,
\end{equation}
with the quartic
\begin{align}
\label{eq:kummer-quartic}
K=K(x,y,z,t)&=x^4+y^4+z^4+t^4+2Dxyzt \notag \\
&\phantom{{}={}}+A(x^2t^2+y^2z^2)+B(y^2t^2+x^2z^2) \notag \\
&\phantom{{}={}}+C(z^2t^2+x^2y^2)\,.
\end{align}
\end{subequations}
The parameters $A\dots D$ must be chosen as
\begin{subequations}
\begin{align}
A&=2[1+8(k^1)^2]\,,\quad B=C=2[1+2(k^1)^2]\,, \\[1ex]
D&=-16\mathrm{i}(k^1)^3\,.
\end{align}
\end{subequations}
for $k^1$, and for $k^2$ we arrive at
\begin{subequations}
\begin{align}
A&=C=2[1+2(k^2)^2]\,,\quad B=2[1+8(k^2)^2]\,, \\[1ex]
D&=-16\mathrm{i}(k^2)^3\,.
\end{align}
\end{subequations}
In fact, a quartic surface $K=0$ with $K$ given by Eq.~\eqref{eq:kummer-quartic} is known as Kummer's surface in the mathematics literature; see, e.g., p.~81 in Ref.~\cite{Hudson:1905} and Ref.~\cite{Cayley:1846} for additional information. Note that the dispersion equation is expressed in terms of a complexified version of $K$, which is a consequence of the Lorentzian metric signature. In fact, Kummer's surface is a generalization of Fresnel's wave surface of Eq.~\eqref{eq:dispersion-equation-k34567-transformed} and has more intricate properties.
\begin{figure}
\centering
\includegraphics[scale=0.25]{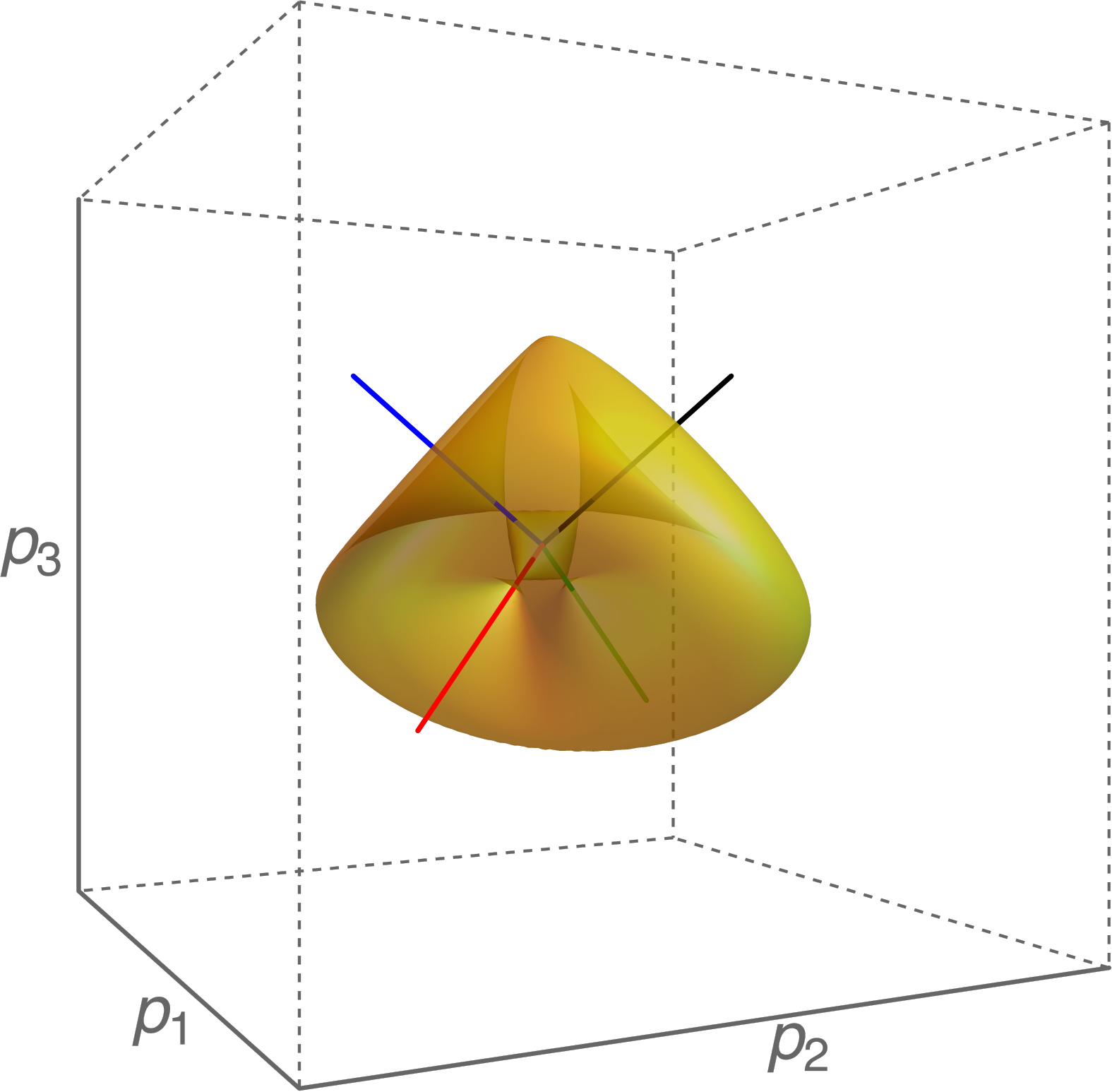}
\caption{Surface of constant $p^0$ from Eq.~\eqref{eq:dispersion-equation-k12}. The presence of two shells as well as four singular points can be noted. However, there are no optical axes in Fresnel's sense, but four distinct directions pointing from the center to the singular points each.}
\label{fig:surface-k1}
\end{figure}

Plotting Eq.~\eqref{eq:dispersion-equation-k12} for constant $p^0$ again provides a closed surface with outer and inner shells; see Fig.~\ref{fig:surface-k1}. There are four singular points, where both shells intersect with each other. Differently from Fresnel's wave surface, the four wave vectors associated with the singular points do not lie in a single plane. For $k^1$, the four directions are given by
\begin{subequations}
\label{eq:directions-surface-kummer}
\begin{align}
\mathbf{p}_{1,2}&=\begin{pmatrix}
f(k^1) \\
\pm 1 \\
\pm 1 \\
\end{pmatrix}\,,\quad \mathbf{p}_{3,4}=\begin{pmatrix}
-f(k^1) \\
\pm 1 \\
\mp 1 \\
\end{pmatrix}\,, \\[1ex]
f(x)&=\sqrt{\sqrt{(1+x^2)(1+4x^2)}-(1+2x^2)}\,.
\end{align}
\end{subequations}
For $k^2$, the first two components of these spatial vectors must be switched. Since all four vectors are linearly independent, pairs of these vectors can be formed such that each pair lies in a single plane. Thus, there are two distinct planes. Because of this, the concept of optical axes loses its meaning for $k^1$ and $k^2$. Instead, we encounter a new type of birefringence that goes beyond uniaxial and biaxial scenarios usually studied in the literature. Note that the limits $k^{1,2}\rightarrow 0$ recover biaxial media with optical axes corresponding to those for the coefficients $k^3$ and $k^4$, respectively; cf.~Tab.~\ref{tab:vectors-dispersions-sector-2}.

\begin{figure}
\centering
\includegraphics[scale=0.25]{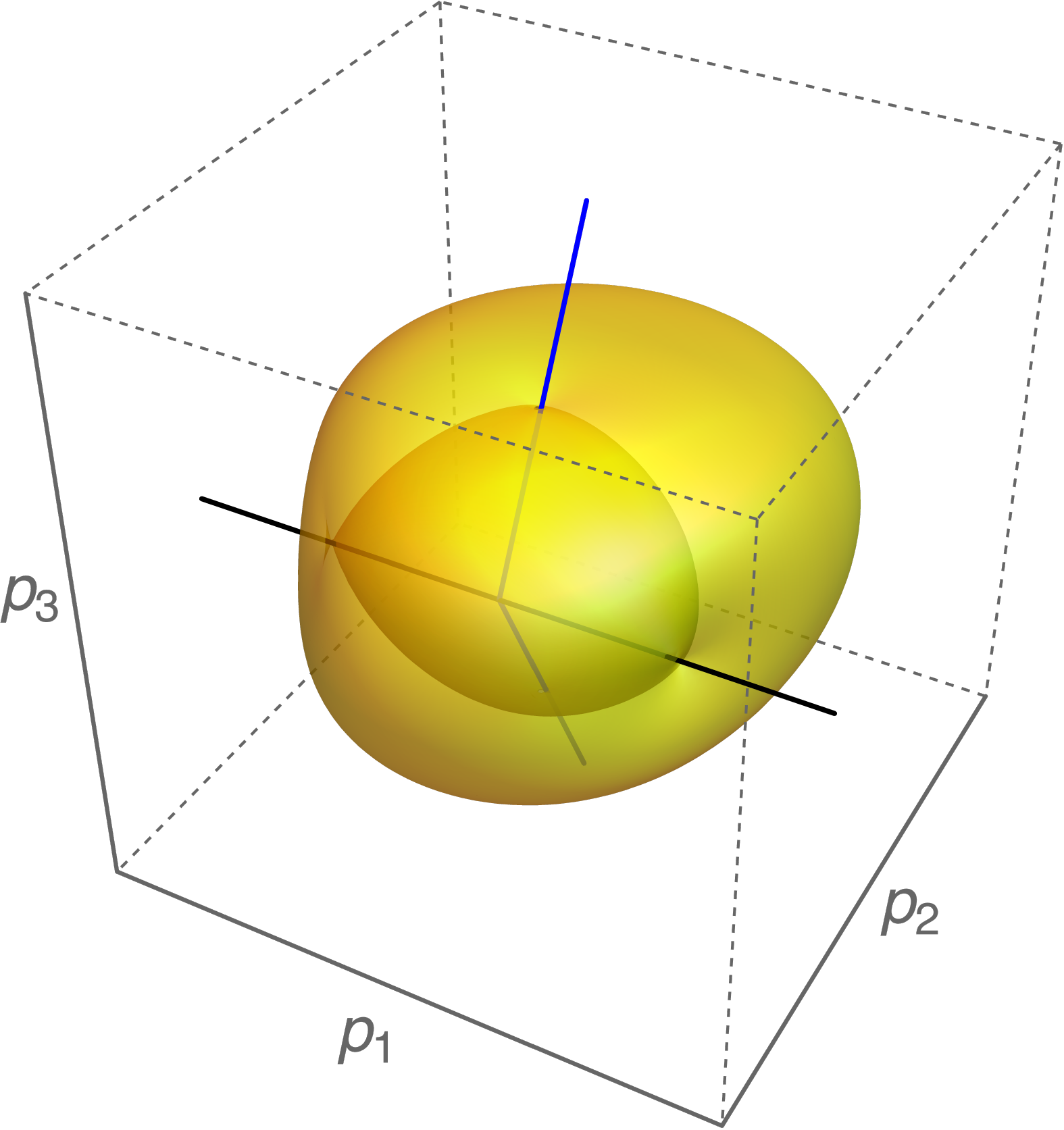}
\caption{Surface of constant $p^0$ from Eq.~\eqref{eq:dispersion-equation-k8910}. The single optical axis, which connects two singular points, is illustrated as a black line. The directions pointing to the remaining singular points are shown in blue.}
\label{fig:surface-k8}
\end{figure}

Finally, the dispersion relations for the coefficients $k^8\dots k^{10}$ are presumably the most complicated ones. They can be expressed in the form
\begin{equation}
\label{eq:dispersion-equation-k8910}
0=K(p^1,p^2,p^3,\mathrm{i}p^0)+\Delta K^{(a)}(p^1,p^2,p^3,\mathrm{i}p^0)\,,
\end{equation}
with the quartic of Eq.~\eqref{eq:kummer-quartic} for $A=B=C=2$ and $D=0$ and additional contributions $\Delta^{(a)}K=\Delta^{(a)}K(x,y,z,t)$, which are explicitly given by
\begin{subequations}
\begin{align}
\Delta K^{(8)}&=4(k^8)^2z^4+4\mathrm{i}k^8yt(x^2+y^2+z^2+t^2) \notag \\
&\phantom{{}={}}+4(k^8)^2(y^2+t^2)z^2\,, \displaybreak[0]\\[2ex]
\Delta K^{(9)}&=4(k^9)^2y^4+4\mathrm{i}k^9zt(x^2+y^2+z^2+t^2) \notag \\
&\phantom{{}={}}+4(k^9)^2(z^2+t^2)y^2\,, \displaybreak[0]\\[2ex]
\Delta K^{(10)}&=4(k^{10})^2z^4-4\mathrm{i}k^{10}xt(x^2+y^2+z^2+t^2) \notag \\
&\phantom{{}={}}+4(k^{10})^2(x^2+t^2)z^2\,,
\end{align}
\end{subequations}
for $k^8$, $k^9$, and $k^{10}$, respectively. According to Ref.~\cite{Baekler:2014kha}, these dispersion equations must be Kummer surfaces, although they do not share the explicit form of Eq.~\eqref{eq:dispersion-equation-k12} with $k^1$ and $k^2$. Plotting each dispersion equation for constant $p^0$ provides two closed shells intersecting each other at four singular points; see Fig.~\ref{fig:surface-k8}. The associated wave vectors again do not lie in a single plane, unlike in the cases $k^3\dots k^7$.

However, unlike for the coefficients $k^1$ and $k^2$, two wave vectors are antiparallel, which leads us to conclude that there is a single optical axis for each coefficient. The latter is $\mathbf{e}_x$ for $k^{8,9}$ and $\mathbf{e}_y$ for $k^{10}$. Nevertheless, these are not simple uniaxial cases like those of the first principal sector of $k_F$; see Tab.~\ref{tab:birefringence-sector-1}. The wave vectors pointing towards the other two singular points of the dispersion surface are
\begin{subequations}
\label{eq:directions-surface-k8k9k10}
\begin{align}
\mathbf{p}_{1,2}&=\begin{pmatrix}
0 \\
g(k^8) \\
\pm 1 \\
\end{pmatrix}\,,\quad\mathbf{p}_{1,2}=\begin{pmatrix}
0 \\
\pm 1 \\
g(k^9) \\
\end{pmatrix}\,, \\[1ex]
\mathbf{p}_{1,2}&=\begin{pmatrix}
g(k^{10}) \\
0 \\
\pm 1 \\
\end{pmatrix}\,,
\end{align}
for $k^8$, $k^9$, and $k^{10}$, respectively, where
\begin{equation}
g(x)=\frac{1}{\sqrt{2}}\sqrt{\sqrt{1+4x^2}-1}\,.
\end{equation}
\end{subequations}
In principle, pairs of wave vectors can be put into distinct planes, where these planes intersect each other along the optical axes shared by two of the four vectors. Note that the limits $k^{8,9,10}\rightarrow 0$ reproduce the biaxial cases of the coefficients $k^6$, $k^5$, and $k^7$, respectively; cf.~Tab.~\ref{tab:vectors-dispersions-sector-2}.

\subsubsection{Dispersion relations}

The dispersion relations for $k^3\dots k^7$ can be conveniently expressed via spatial vectors $\mathbf{u}$ and $\mathbf{v}$ as follows:
\begin{align}
\label{eq:dispersions-birefringence-sector-2}
\omega^{(\pm)}(\mathbf{p})&=\bigg[\mathbf{p}^2+2k^a(\mathbf{u}\cdot\mathbf{p})(\mathbf{v}\cdot\mathbf{p}) \notag \\
&\phantom{{}={}}\,\pm 2k^a\sqrt{(\mathbf{p}\cdot P_u\cdot \mathbf{p})(\mathbf{p}\cdot P_v\cdot \mathbf{p})}\,\bigg]^{\frac{1}{2}}\,,
\end{align}
which holds for the dispersion equation in the form of Eq.~\eqref{eq:dispersion-equation-k34567}, i.e., before carrying out any coordinate transformation.
Here, $a\in\{3\dots 7\}$ and the result is expressed in terms of projectors $P_u:=\mathds{1}_3-\mathbf{u}\mathbf{u}^T$ and $P_v:=\mathds{1}_3-\mathbf{v}\mathbf{v}^T$.
Table~\ref{tab:vectors-dispersions-sector-2} provides the vectors $\mathbf{u}$ and $\mathbf{v}$ for the SME coefficients whose dispersion relations can be written according to Eq.~\eqref{eq:dispersions-birefringence-sector-2}. Note that there is a correspondence between the four-vectors of Tab.~\ref{tab:parametrization-second-principal-sector} for the parametrization of $k_F$ and the spatial vectors $\mathbf{u},\mathbf{v}$ in Eq.~\eqref{eq:dispersions-birefringence-sector-2}, chosen according to Tab.~\ref{tab:vectors-dispersions-sector-2}.

Several comments are in order with regards to Tab.~\ref{tab:parametrization-second-principal-sector}. First, for configurations expressed in terms of purely spacelike $\xi^{(A)},\xi^{(B)}$ the vectors $\mathbf{u},\mathbf{v}$ correspond to their spacelike parts, i.e., $\mathbf{u}=\boldsymbol{\xi}^{(A)}$ and $\mathbf{v}=\boldsymbol{\xi}^{(B)}$. Second, when the configurations involve bilinear combinations of four-vectors $\xi^{(A)},\xi^{(B)}$ with themselves, such as for $k^3$ and $k^4$, the spatial vectors $\mathbf{u},\mathbf{v}$ are linear combinations of the spacelike parts $\boldsymbol{\xi}^{(A)},\boldsymbol{\xi}^{(B)}$. Finally, when the purely timelike vector $\xi^{(0)}$ occurs, the corresponding dispersion relations involve cubic roots and cannot be written in the form of Eq.~\eqref{eq:dispersions-birefringence-sector-2}.

The latter form of the dispersion relations directly reveals the optical axes for those cases. Their directions must satisfy
\begin{equation}
(\mathbf{p}^2-(\mathbf{u}\cdot\mathbf{p})^2)(\mathbf{p}^2-(\mathbf{v}\cdot\mathbf{p})^2)=0\,.
\end{equation}
This condition is immediately fulfilled by either $\mathbf{p}=\mathbf{u}$ or $\mathbf{p}=\mathbf{v}$, i.e., $\mathbf{u}$ and $\mathbf{v}$ are interpreted as optical axes. Settings with $\mathbf{u}\neq \mathbf{v}$ describe biaxial birefringence, which is the case for the five coefficients listed in Tab.~\ref{tab:vectors-dispersions-sector-2}.
\begin{table}
\begin{tabular}{cccc}
\toprule
Coefficient $k^a$ & Type & \multicolumn{2}{c}{Optical axes} \\
\cmidrule(lr){3-4}
                &      & $\mathbf{u}$ & $\mathbf{v}$ \\
\midrule
$k^3$ & Biaxial & $(\mathbf{e}_y-\mathbf{e}_x)/\sqrt{2}$ & $(\mathbf{e}_y+\mathbf{e}_x)/\sqrt{2}$ \\
$k^4$ & Biaxial & $(\mathbf{e}_z-\mathbf{e}_x)/\sqrt{2}$ & $(\mathbf{e}_x+\mathbf{e}_z)/\sqrt{2}$ \\
$k^5$ & Biaxial & $\mathbf{e}_x$ & $\mathbf{e}_y$ \\
$k^6$ & Biaxial & $\mathbf{e}_x$ & $\mathbf{e}_z$ \\
$k^7$ & Biaxial & $\mathbf{e}_y$ & $\mathbf{e}_z$ \\
\bottomrule
\end{tabular}
\caption{Vectors $\mathbf{u}$ and $\mathbf{v}$ for the SME coefficients $k^a$ whose dispersion relations can be expressed in the form of Eq.~\eqref{eq:dispersions-birefringence-sector-2}. The first column provides the SME coefficient and the second lists the birefringence type. The third and fourth columns state the optical axes.}
\label{tab:vectors-dispersions-sector-2}
\end{table}
It is possible to consider several of the coefficients of Tab.~\ref{tab:vectors-dispersions-sector-2} nonzero at the same time. The generic dispersion relation is then of the form
\begin{align}
\label{eq:dispersions-birefringence-sector-2-generic}
\omega^{(\pm)}&=\bigg[\mathbf{p}^2+2\sum_a k^a\,(\mathbf{u}_a\cdot\mathbf{p})(\mathbf{v}_a\cdot\mathbf{p}) \notag \\
&\phantom{{}={}}\,\pm 2\sqrt{\sum_a (k^a)^2(\mathbf{p}\cdot P_{u_a}\cdot \mathbf{p})(\mathbf{p}\cdot P_{v_a}\cdot \mathbf{p})+\text{cpl.}}\,\bigg]^{\frac{1}{2}},
\end{align}
where $a\in\{3\dots 7\}$ and ``cpl.'' stands for coupling terms between distinct coefficients. The latter have different structures depending on the coefficients considered. For example, for $k^5$ and $k^6$, one finds
\begin{align}
\text{cpl.}&\supset k^5k^6(\mathbf{v}_5\cdot \mathbf{p})(\mathbf{v}_6\cdot\mathbf{p})\Big[-(\mathbf{u}_5\cdot\mathbf{u}_6+\mathbf{v}_5\cdot\mathbf{v}_6)\mathbf{p}^2 \notag \\
&\hspace{3.6cm}+(\mathbf{u}_5\cdot\mathbf{p})(\mathbf{u}_6\cdot\mathbf{p})\Big]\,.
\end{align}
The dispersion relations for $k^1$ and $k^2$, which follow from Eq.~\eqref{eq:dispersion-equation-k12}, are highly complicated and depend on cubic roots, which is why they will not be stated here. However, they are quite simple for wave vectors along the directions of Eq.~\eqref{eq:directions-surface-kummer}. Then, there is a single relatively compact dispersion relation of the form
\begin{subequations}
\begin{align}
\omega(p)&=\Xi(k^{1,2}) p\,, \\[1ex]
\Xi(x)&=\frac{1}{3}\left[\sqrt{1+4x^2}+2\sqrt{1+x^2}\,\right]\,,
\end{align}
\end{subequations}
where $p$ is the magnitude of the wave vector along such a direction. Therefore, the refractive index along these directions simply amounts to
\begin{equation}
n=\frac{p}{\omega(p)}=\frac{1}{\Xi(k^{1,2})}\,,
\end{equation}
which is the same for each direction and can be interpreted as an experimental signature of such configurations. The dispersion relations in planes orthogonal to each of Eq.~\eqref{eq:directions-surface-kummer} are anisotropic and have an intricate form.

The dispersion relations for the remaining coefficients $k^8\dots k^{10}$ follow from Eq.~\eqref{eq:dispersion-equation-k8910}. They also depend on cubic roots and are therefore omitted. However, along the directions of Eq.~\eqref{eq:directions-surface-k8k9k10}, they simply collapse to the vacuum result $\omega(p)=p$, such that the refractive index is, indeed, $n=1$ for wave propagation along these directions. As for $k^1$ and $k^2$, the dispersion relations are anisotropic and complicated in planes orthogonal to each of Eq.~\eqref{eq:directions-surface-k8k9k10}.

\subsection{Carroll-Field-Jackiw theory}

Electromagnetic-wave propagation in optical media for arbitrarily large $(k_{AF})^{\mu}$ coefficients and higher-derivative extensions of CFJ theory was studied in Ref.~\cite{Silva:2021fzh}. The dispersion equation of CFJ theory is well-known and was already obtained in Ref.~\cite{Carroll:1989vb}. It makes sense to distinguish between a purely timelike and a purely spacelike $k_{AF}$, respectively. In these cases, the dispersion relations are expressed in terms of square roots. For the purely timelike sector, the quartic equation factorizes, which provides the simple result
\begin{equation}
\omega^{(\pm)}=\sqrt{\mathbf{p}^2\pm |(k_{AF})^0||\mathbf{p}|}\,.
\end{equation}
Note that $p_0^{(-)}$ takes complex values for $|\mathbf{p}|<|(k_{AF})^0|$, which implies either unitarity or microcausality problems, as shown in Ref.~\cite{Adam:2001kx}. This issue always occurs independently of the magnitude of $|k_{AF}^0|$, which renders the physical significance of the timelike sector of CFJ theory questionable.

The purely spacelike sector exhibits the dispersion relations
\begin{equation}
\omega^{(\pm)}=\sqrt{\mathbf{p}^2+\frac{(\mathbf{k}_{AF})^2}{2}\pm\sqrt{\frac{(\mathbf{k}_{AF})^4}{4}+(\mathbf{k}_{AF}\cdot\mathbf{p})^2}}\,,
\end{equation}
expressed in terms of the spacelike part $\mathbf{k}_{AF}$ of the four-vector $k_{AF}$. Here, $\omega^{(+)}$ is manifestly real, and $\omega^{(-)}$ can be checked explicitly to be real as well. Thus, the spacelike sector is not plagued by the same problems as the timelike sector.

Note that both sectors do not exhibit optical axes, since the dispersion surfaces do not intersect each other. Thus, there are no singular points. The nature of the CFJ term is profoundly different from that of the modified Maxwell term because of the presence of a mass scale $[(k_{AF})^{\mu}]=1$. The latter separates the two shells from each other. Recall that the CFJ term is equivalent to an inhomogeneous $\theta$ term up to a surface contribution; see Eq.~\eqref{eq:correspondence-CFJ-theta-theory}.
\begin{figure}
\centering
\subfloat[]{\label{fig:dispersion-CFJ-timelike}\includegraphics[scale=0.3]{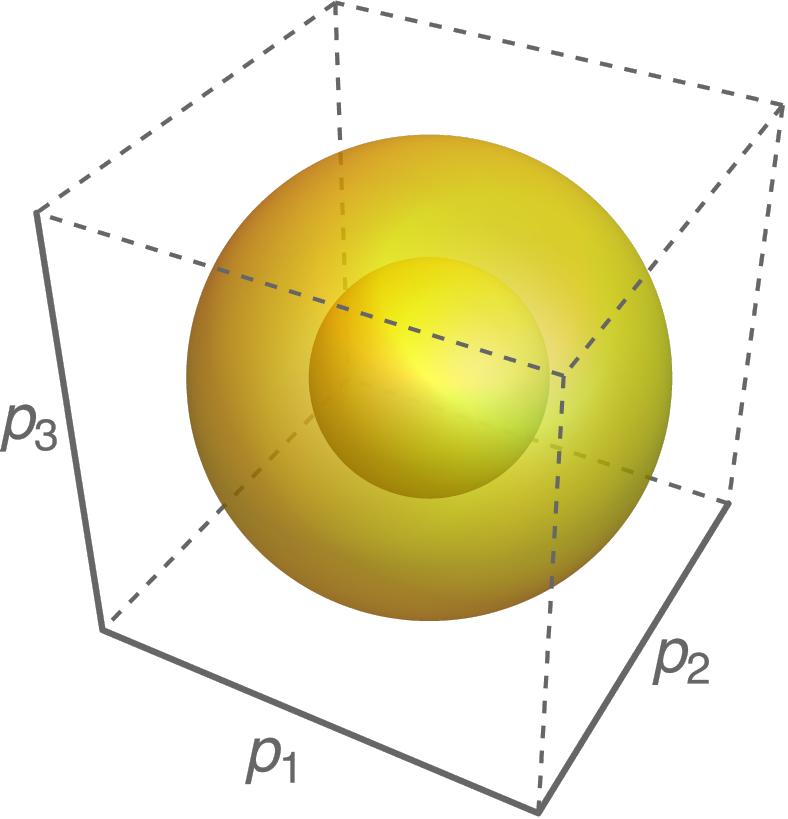}}\quad
\subfloat[]{\label{fig:dispersion-CFJ-spacelike}\includegraphics[scale=0.3]{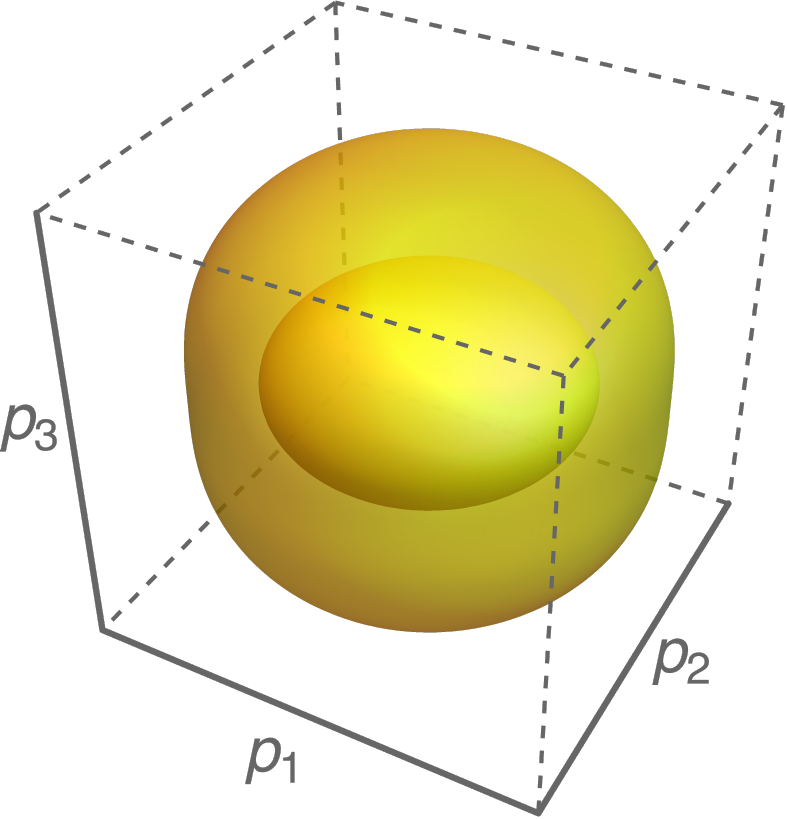}}
\caption{Dispersion surfaces for the purely timelike \protect\subref{fig:dispersion-CFJ-timelike} and purely spacelike regime \protect\subref{fig:dispersion-CFJ-spacelike}, respectively.}
\end{figure}

As discussed previously, for Weyl semimetals, it holds that $\theta(x)\sim b\cdot x$ with the vector-valued background field $b_{\mu}$ of the SME fermion sector. Weyl semimetal phases have been identified experimentally~\cite{Lv:2021} with timelike components $|b^0|\approx \unit[100]{meV}$ and $|\mathbf{b}|\approx \unit[100]{eV}$, respectively. The timelike components $b^0$ and $(k_{AF})^0$ violate P and preserve T, whereas the spacelike components $\mathbf{b}$ and $\mathbf{k}_{AF}$ violate T and preserve P. 

Note that the most prominent example, TaAs, is a nonmagnetic and noncentrosymmetric material that exhibits P violation but preserves T. The latter is described by a nonzero $\mathbf{b}$, which corresponds to a nonzero $\mathbf{k}_{AF}$ of the CFJ term. The magnetic point group of TaAs is \textbf{4mm}, i.e., according to Tab.~\ref{tab:magnetoelectric}, a $\theta$ term is not compatible with this material. However, one must also take into account the transformation behavior of $x^{\mu}$ under P and T. This example shows an additional caveat in the analysis of electromagnetic properties based on crystallographic groups and magnetic point groups when coefficients exhibit an additional spacetime dependence.

\subsection{Material media}

As we have already noted, the coefficients of the second principal sector are suppressed for materials governed by the point groups \textbf{2/m}, \textbf{222}, \textbf{mm2}, and \textbf{mmm}; see Tab.~\ref{tab:example-materials}. Thus, birefringence of such materials is presumably almost completely governed by the first principal sector. For increasing symmetry, $k^a$ may contribute, as is the case for \textbf{32}. However, the relationship $k^4=-2k^3$ turns the material into a uniaxial one, contrary to what is expected from $k^a$. Hence, the presence of nonzero coefficients of the second principal sector does not automatically imply that birefringence is biaxial.

Magnetoelectric materials have more complicated properties, which is why more nonzero SME coefficients are needed to parametrize them. For materials of the magnetic point group $\bar{\textbf{3}}$'\textbf{m}', nonzero coefficients $k^{1,2}$ emerge that have not been observed for the other materials considered in Tab.~\ref{tab:example-materials}. However, these coefficients are again related via $k^2=-k^1$. Therefore, exotic birefringence effects that we witnessed for $k^{1,2}$ via Eq.~\eqref{eq:dispersion-equation-k12} do not occur for this material. Instead, it is simply uniaxial. This again emphasizes that care must be taken when predicting effects for several nonzero $k^a$ coefficients that depend on each other.

The examples that we looked at previously allow for a certain conclusion. The crystallographic and magnetic point groups describing crystal lattices either eliminate SME coefficients or imply relationships between several of them. Consequently, any possibility of exotic birefringence properties not studied in detail in the literature is likely to be excluded for natural crystals. However, these restrictions are not expected to have significance for artificial states of matter such as metamaterials~\cite{Sun:2025}.

Having said this, we predict that artificial materials exist whose magnetoelectric properties are described by coefficients of the sets $\{k^1,k^2\}$ and $\{k^8\dots k^{10}\}$ without any restrictive relationships imposed between individual coefficients. We challenge materials scientists to conceive ideas for artificial materials with these magnetoelectric couplings. According to the findings presented here, such materials are expected to have characteristics not found in natural optical media. Recall that the $\theta$ term of Eq.~\eqref{eq:lagrangian-theta}, which originally found its way into high-energy physics as a purely hypothetical proposal \cite{Callan:1976je,Wilczek:1987mv}, has been observed in material media~\cite{Sihvola:1995,Hehl:2007jy,Hehl:2007ut}. Thus, it is plausible that extensions of the $\theta$ term also play a role in materials science.

\subsection{Comparison to studies beyond the SME}

As mentioned at the beginning of this paper, the electromagnetism of unusual materials with nontrivial permeability or magnetoelectric couplings has been delved into in Refs.~\cite{Obukhov:2000nw,Obukhov:2004zz,Hehl:2007jy,Hehl:2007ut,Baekler:2014kha,Favaro:2014lja,Favaro:2015jxa,Hehl:2016wwp}. By and large, our observations are in accordance with those made beyond the SME framework. The authors of the previous works define an electromagnetic response tensor in analogy to ours of Eq.~\eqref{eq:electromagnetic-response}, modulo irrelevant factors.
\begin{figure}
\centering
\includegraphics[scale=0.25]{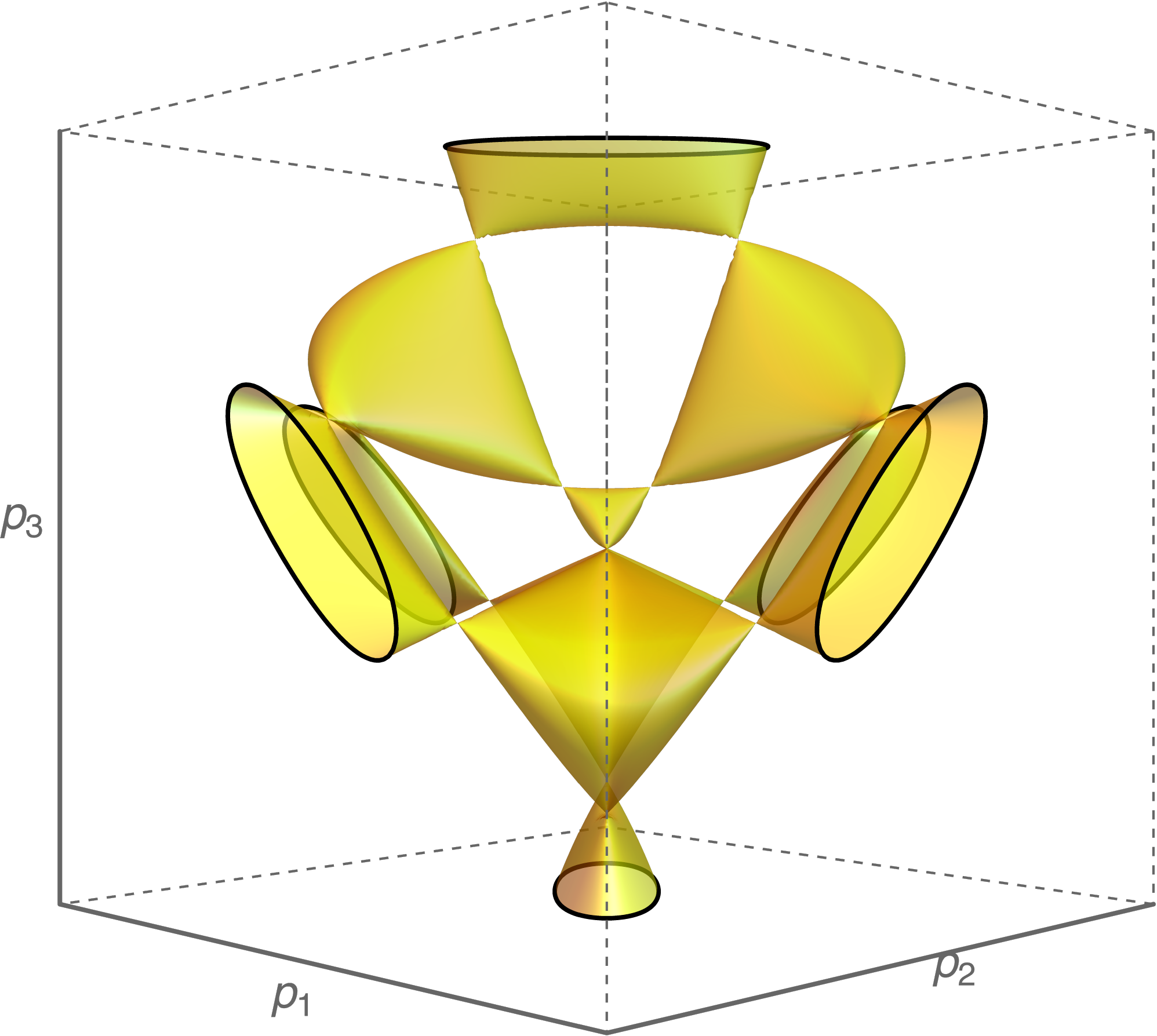}
\caption{Surface of constant $p^0$ based on Eq.~\eqref{eq:case-3-hehl}.}
\label{fig:surface-hehl}
\end{figure}

They then classify their electrodynamics in material media according to a decomposition of their response tensor into pieces that transform under irreducible representations of the Lorentz group~\cite{Obukhov:2004zz}. Their first piece involves 20 coefficients and corresponds to the $k_F$ sector of the minimal SME, without imposing the double tracelessness of~$k_F$. The second sector is based on an electromagnetic response tensor that is devoid of the symmetry with respect to interchanging neighboring blocks. They call this piece the skewon. The latter is not contained in the SME, since the skewon cannot arise from an action, as the authors remark. The third piece contains a single pseudoscalar coefficient corresponding to the $\theta$ term outlined in Sec.~\ref{eq:CFJ-theory}.

The authors of the previous papers also delve into dispersion equations for different choices of the tensors $\epsilon$, $\mu$, and $\alpha$. They identify sectors described by Fresnel's wave surface, as is the case for the SME coefficients $k^3\dots k^7$. Configurations associated with quartic surfaces different from Fresnel's are also found~\cite{Baekler:2014kha,Favaro:2015jxa}. As an example, we quote the third case of the latter paper that translates to the SME as follows:
\begin{subequations}
\label{eq:case-3-hehl}
\begin{align}
\tilde{\kappa}^{00}&=-\frac{3}{4}\,,\quad \tilde{\kappa}^{11}=\tilde{\kappa}^{22}=\frac{1}{4}(\sqrt{3}-2)\,, \\[1ex]
k^1&=k^2=\frac{1}{8}(\sqrt{3}+3)\,,\quad k_{\mathrm{tr}}=-6\,.
\end{align}
\end{subequations}
This case has an intriguing dispersion surface that is different from those illustrated previously, which is why we reproduce it in Fig.~\ref{fig:surface-hehl}. Note that this surface is not compact.

Based on algebraic methods, the authors of Ref.~\cite{Baekler:2014kha} prove that Kummer's quartic surface describes all possibilities, as long as the skewon is not taken into account. The fate of the skewon part remains unclear. It is conjectured that its dispersion equations require generalizations of Kummer's quartic surface~\cite{Obukhov:2004zz,Favaro:2014lja}, but a final decisive statement seems elusive to this day.

\section{Conclusions and outlook}
\label{eq:conclusions}

This paper was compiled to raise awareness of the possibilities that the SME provides for describing electromagnetic phenomena in material media. Years ago, some authors already emphasized that the covariant formalism of electrodynamics had been gaining traction among practitioners and engineers~\cite{Hehl:2016wwp}. After all, electromagnetism is a relativistic \textit{U}(1) gauge theory, and certain properties only become transparent in a relativistic approach. We have shown that the electromagnetic sector of the SME is versatile in its application to material media. We particularly focused on the relationship between crystallographic groups of atomic lattices and configurations of SME coefficients governing the electromagnetic properties of materials at the macroscopic level. The SME permits a systematic parametrization of such properties. Hence, the findings presented here have the potential to contribute to the design of new optical media.

A particular interest of ours was to look into the nature of birefringence associated with each SME coefficient. We encountered effects that have not been thoroughly addressed in the modern literature. The most interesting coefficients are contained in the sets $\{k^1,k^2\}$ and $\{k^8\dots k^{10}\}$ and play a role in magnetoelectric couplings. These are capable of inducing birefringence that is beyond the common description via optical axes. The wave surfaces have forms that differ significantly from Fresnel's surface usually considered. These results shall serve as inspiration for designing optical media with such unusual properties.

We are aware that the design of such materials is likely to be challenging. Recall the intriguing material $\mathrm{Cr_2O_3}$ that was found to exhibit a magnetoelectric coupling involving a pseudoscalar coefficient~\cite{Sihvola:1995,Hehl:2007jy,Hehl:2007ut}. The authors of the first two papers beg the question of whether it would be experimentally feasible to design a material whose magnetoelectric coupling is a pure pseudoscalar. To the best of our knowledge, this goal has not been achieved until now.

A material with vanishing permittivity and permeability, but nonzero magnetoelectric couplings parametrized by the above SME coefficients is expected to have intriguing optical properties. The only coefficient compatible with $\epsilon_{ij}=\mu_{ij}=0$ is $k_{\mathrm{tr}}=-6$. Since the latter is not associated with any particular crystal structure, we conclude that such a material is not expected to occur in nature. However, the reader may recall metamaterials~\cite{Sun:2025}, which are completely artificial and exhibit negative refractive indices at an effective level. Materials with the properties sought could be conceived by related means. Note that, for example, effects predicted for Tellegen media have been identified in metamaterials recently~\cite{Yang:2025}.

The main advantages of resorting to the SME are its action-based approach and generality. Thus, we make the bold prediction that materials, possibly artificial ones, should exist such that each single SME coefficient plays a role in parametrizing their electromagnetic properties. In particular, this should apply to $k^1,k^2$ and $k^8\dots k^{10}$ of the $k_F$ tensor, which are linked to highly unusual characteristics that are mostly unexplored in the literature. Our conviction is that such exotic optical media probably do not simply exist as natural compounds. Therefore, we propose that materials scientists examine the feasibility of designing materials of this kind.

Another interesting direction is to assess possible extensions of bi-isotropic materials, such as those investigated in Refs.~\cite{Silva:2022sps,Silva:2022bnv}. Maybe it is feasible to design materials that are tri-isotropic such that permittivity, permeability, and magnetoelectric couplings are nontrivial at the same time. Although each of these proposals may be a challenging endeavor, the materials conceived are expected to have extraordinary properties not found in optical media that are abundant in nature. The authors of Ref.~\cite{Sun:2025} emphasize how the design of metamaterials is an art in its own right, which breaks through the paradigms of materials science based on conventional states of matter.

From the theory side, various systematic extensions of the findings made are expected to be possible. First, higher-derivative operators, as contained in the nonminimal SME~\cite{Kostelecky:2009zp,Kostelecky:2011gq,Kostelecky:2013rta,Ding:2016lwt,Kostelecky:2018yfa,Kostelecky:2020hbb}, can incorporate dispersion and other effects. In materials science, such extensions have been proposed for, e.g., Tellegen media~\cite{Barredo-Alamilla:2023lpu,Barredo-Alamilla:2024kjy}. Second, thermal field theory should allow for including temperature effects properly. Third, since the SME is based on an action, quantum effects arise naturally and can be treated via perturbation theory.

Note that the SME even goes beyond electromagnetism. It parametrizes modifications of the Dirac fermion sector as well as the non-Abelian gauge theories of the electroweak and strong interactions. Due to its generality, we expect that it will predict even novel material effects, only to be found in, e.g., the non-Abelian sectors. The SME fermion sector was shown to house descriptions of both Weyl and Dirac semimetals~\cite{Kostelecky:2021bsb}. The existence of nodal lines of dispersion surfaces in momentum space, as well as drumhead surface states, can be attributed to a specific class of SME coefficients known as $g$. Recent research demonstrates how geometrical phases and topological indices arise in such settings and what we are able to learn from them~\cite{Kostelecky:2025zsy}. Hence, the SME provides a vast number of innovative vistas in different areas of materials science.

\section*{Acknowledgments}

It is a pleasure to thank F.W.~Hehl for several valuable discussions on electrodynamics in materials and exotic dispersion equations. MS is indebted to CNPq Produtividade 307653/2025-0 and CAPES/Finance Code 001.

\appendix

\section{Auxiliary mathematical relationships}
\label{app:formulas}

When dealing with symmetry transformations in crystals, the following form of the rotation matrix describing a rotation with angle $\Omega$ around the rotation axis indicated by the unit vector $\hat{\boldsymbol{\omega}}$ is valuable:
\begin{subequations}
\label{eq:generic-rotation}
\begin{align}
\mathcal{R}^{[\hat{\boldsymbol{\omega}}]}(\Omega)&=\hat{\boldsymbol{\omega}}\hat{\boldsymbol{\omega}}^T+(\mathds{1}_3-\hat{\boldsymbol{\omega}}\hat{\boldsymbol{\omega}}^T)\cos(\Omega) \notag \\
&\phantom{{}={}}+\begin{pmatrix}
0 & -\hat{\omega}_3 & \hat{\omega}_2 \\
\hat{\omega}_3 & 0 & -\hat{\omega}_1 \\
-\hat{\omega}_2 & \hat{\omega}_1 & 0 \\
\end{pmatrix}\sin(\Omega)\,, \\[2ex]
\hat{\omega}_3&=\sqrt{1-(\hat{\omega}_1)^2-(\hat{\omega}_2)^2}\,.
\end{align}
\end{subequations}
The dispersion equation~\eqref{eq:dispersion-equation-sector-1}, which is expressed in terms of the wave four-vector $p_{\mu}$, can be brought into the form of a fourth-order polynomial equation in terms of $p_0$.
The result is lengthy but is printed here for further use:
\begin{widetext}
\begin{subequations}
\label{eq:dispersion-polynomial-sector-1}
\begin{align}
0&=\rho p_0^4+\sigma p_0^3+\tau p_0^2+\upsilon p_0+\chi\,, \\[2ex]
\rho&=1-\frac{1}{2}\tilde{\kappa}_{\mu\nu}\tilde{\kappa}^{\mu\nu}+\frac{1}{3}\tilde{\kappa}^{\mu}_{\phantom{\mu}\nu}\tilde{\kappa}^{\nu}_{\phantom{\nu}\varrho}\tilde{\kappa}^{\varrho}_{\phantom{\varrho}\mu}+\tilde{\kappa}^{0\beta}\tilde{\kappa}_{\beta}^{\phantom{\beta} 0}+\tilde{\kappa}^{00}(2+\tilde{\kappa}^{00}+\tilde{\kappa}^{0\beta}\tilde{\kappa}_{\beta}^{\phantom{\beta} 0})-\tilde{\kappa}^{0\beta}\tilde{\kappa}_{\beta\gamma}\tilde{\kappa}^{\gamma 0}\,, \displaybreak[0]\\[2ex]
\sigma&=-2\Big[(1+\tilde{\kappa}^{00})(2\tilde{\kappa}^{0i}p^i+p^i\tilde{\kappa}^{i\beta}\tilde{\kappa}_{\beta} ^{\phantom{\beta}0})+\tilde{\kappa}^{0i}p^i\tilde{\kappa}^{0\beta}\tilde{\kappa}_{\beta}^{\phantom{\beta}0}-\tilde{\kappa}^{0\beta}\tilde{\kappa}_{\beta\gamma}\tilde{\kappa}^{\gamma i}p^i\Big]\,, \displaybreak[0]\\[2ex]
\tau&=2\left(1+\tilde{\kappa}^{00}+\frac{1}{2}\tilde{\kappa}^{0\beta}\tilde{\kappa}_{\beta}^{\phantom{\beta}0}\right)p^i\tilde{\kappa}^{ij}p^j+\bigg[-2\left(1+\tilde{\kappa}^{00}-\frac{1}{2}\tilde{\kappa}_{\mu\nu}\tilde{\kappa}^{\mu\nu}+\frac{1}{3}\tilde{\kappa}^{\mu}_{\phantom{\mu}\nu}\tilde{\kappa}^{\nu}_{\phantom{\nu}\varrho}\tilde{\kappa}^{\varrho}_{\phantom{\varrho}\mu}\right)-\tilde{\kappa}^{0\beta}\tilde{\kappa}_{\beta}^{\phantom{\beta}0}\tilde{\kappa}^{0\beta}\tilde{\kappa}_{\beta\gamma}\tilde{\kappa}^{\gamma 0}\bigg]\mathbf{p}^2 \notag \\
&\phantom{{}={}}+p^i\tilde{\kappa}^{i\beta}\tilde{\kappa}_{\beta}^{\phantom{\beta}j}p^j-p^i\tilde{\kappa}^{i\beta}\tilde{\kappa}_{\beta}^{\phantom{\beta}\gamma}\tilde{\kappa}^{\gamma j}p^j+4\tilde{\kappa}^{0i}p^i(\tilde{\kappa}^{0j}p^j+p^j\tilde{\kappa}^{j\beta}\tilde{\kappa}_{\beta}^{\phantom{\beta}0})+\tilde{\kappa}^{00}p^i\tilde{\kappa}^{i\beta}\tilde{\kappa}_{\beta}^{\phantom{\beta}j}p^j\,, \displaybreak[0]\\[2ex]
\upsilon&=2\Big[(\tilde{\kappa}^{0i}p^i+p^i\tilde{\kappa}^{i\beta}\tilde{\kappa}_{\beta}^{\phantom{\beta}0}-\tilde{\kappa}^{0\beta}\tilde{\kappa}_{\beta\gamma}\tilde{\kappa}^{\gamma i}p^i)\mathbf{p}^2-p^i\tilde{\kappa}^{ij}p^j(2\tilde{\kappa}^{0k}p^k+p^k\tilde{\kappa}^{k\beta}\tilde{\kappa}_{\beta}^{\phantom{\beta}0})-\tilde{\kappa}^{0i}p^ip^j\tilde{\kappa}^{j\beta}\tilde{\kappa}_{\beta}^{\phantom{\beta}k}p^k\Big]\,, \displaybreak[0]\\[2ex]
\chi&=p^i\tilde{\kappa}^{ij}p^j(p^k\tilde{\kappa}^{kl}p^l+p^k\tilde{\kappa}^{k\beta}\tilde{\kappa}_{\beta}^{\phantom{\beta}l}p^l) \notag \\
&\phantom{{}={}}+\bigg[\left(1-\frac{1}{2}\tilde{\kappa}_{\mu\nu}\tilde{\kappa}^{\mu\nu}+\frac{1}{3}\tilde{\kappa}^{\mu}_{\phantom{\mu}\nu}\tilde{\kappa}^{\nu}_{\phantom{\nu}\varrho}\tilde{\kappa}^{\varrho}_{\phantom{\varrho}\mu}\right)\mathbf{p}^2-2p^i\tilde{\kappa}^{ij}p^j-p^i\tilde{\kappa}^{i\beta}\tilde{\kappa}_{\beta}^{\phantom{\beta}j}p^j+p^i\tilde{\kappa}^{i\beta}\tilde{\kappa}_{\beta\gamma}\tilde{\kappa}^{\gamma j}p^j\bigg]\mathbf{p}^2\,.
\end{align}
\end{subequations}
\end{widetext}

\end{document}